\documentstyle[12pt]{article}

\newcommand{\de}{\partial}
\newcommand{\be}{\begin{equation}}
\newcommand{\ba}{\begin{eqnarray}}
\newcommand{\ea}{\end{eqnarray}}
\newcommand{\ee}{\end{equation}}

\newcommand{\f}{\frac}
\newcommand{\s}{\sqrt}

\newcommand{\ep}{\epsilon}

\newcommand{\beq}{\begin{equation}}
\newcommand{\eeq}{\end{equation}}
\newcommand{\beqa}{\begin{eqnarray}}
\newcommand{\eeqa}{\end{eqnarray}}
\newcommand{\CR}{\nonumber \\}

\newcommand{\th}{\theta}

\newcommand{\bra}[1]{\left\langle\, #1\,\right|}
\newcommand{\ket}[1]{\left|\, #1\,\right\rangle}
\newcommand{\wt}{\widetilde}
\newcommand{\wh}{\widehat}

\newcommand{\ra}{\rightarrow}

\newcommand{\dz}{\mbox{D}0{\rm -}\overline{\mbox{D}0}}

\def\Tr{\mathop{\rm Tr}\nolimits}
\def\TrP{\mathop{\rm Tr\,P}\nolimits}

\def\mat#1{\matt[#1]}
\def\matt[#1,#2,#3,#4]{\left(%
\begin{array}{cc} #1 & #2 \\ #3 & #4 \end{array} \right)}

\begin{document}
\begin{titlepage}
\thispagestyle{empty}
\begin{flushright}
hep-th/0505184 \\
May, 2005 
\end{flushright}

\bigskip

\begin{center}
\noindent{\Large \textbf{
Noncommutativity and Tachyon Condensation}}\\
\vspace{2cm}
\noindent{
Seiji Terashima\footnote{E-mail: seijit@physics.rutgers.edu} 
}\\

\vspace{.2cm}

 {\it  New High Energy Theory Center, Rutgers University\\
126 Frelinghuysen Road, Piscataway, NJ 08854-8019, USA}
\vskip 2em
\bigskip

\end{center}
\begin{abstract}

We study the fuzzy or noncommutative D$p$-branes
in terms of infinitely many unstable D0-branes, from which
we can construct any D$p$-branes.
We show that the tachyon condensation
of the unstable D0-branes
induces the noncommutativity.
In the infinite tachyon condensation limit,
most of the unstable D0-branes disappear 
and remaining D0-branes are actually
the BPS D0-branes with the correct noncommutative 
coordinates.
For the fuzzy $S^2$ case,
we explicitly show only the D0-branes corresponding to
the lowest Landau level survive in the limit.
We also show that
a boundary state for a D$p$-brane 
satisfying the Dirichlet boundary condition on a curved submanifold
embedded in the flat space
is not localized on the submanifold.
This implies that the D$p$-brane on it
is ambiguous at the string scale
and solves the problem 
for a spherical D2-brane with a unit flux on the world volume
which should be equivalent to one D0-brane.
We also discuss the diffeomorphism
in the D0-brane picture.

\end{abstract}
\end{titlepage}

\newpage

\tableofcontents

\section{Introduction}

Noncommutative generalization of 
the geometry is an interesting subject 
partly because it is expected to be related to 
a quantization of the general relativity
which is based on Riemannian geometry.
It has been shown that 
some noncommutative geometry, for examples,
noncommutative torus, noncommutative plane and fuzzy sphere
play important roles
in D-brane physics in string theory \cite{CoDoSc, DoHu, SeWi, Myers}.
It should be emphasized that in these examples
the noncommutative D$p$-brane can be 
equivalently described as BPS D0-branes with
matrix valued coordinate $\Phi^\mu$ which do not commute
each other and this may be the origin of the noncommutativity.

On the other hand,
a noncommutative generalization of the
Riemann geometry was given by Connes
\cite{Connes}.
In this noncommutative geometry,
the spectral triples which
include (a generalization of) the Dirac operator
play a central role.
In string theory, remarkably, 
the spectral triples of
the noncommutative geometry \'{a} la Connes
can be identified 
as configurations of the unstable D0-branes like $\dz$ branes pairs
(or D$(-1)$ branes if we consider Euclidean space-time).
\cite{AsSuTe1,AsSuTe2,AsSuTe3, Te}.
Actually, we can construct 
any D-branes from infinitely many unstable D0-branes.\footnote{
Recently, D-branes are realized as solitons in the gauge theory
with tachyon fields defined on higher dimensional
unstable D-brane systems \cite{Sen0}.
More recently, it was shown that 
D-branes can also be constructed
as bound states of a lower dimensional unstable
D-brane system \cite{Te, AsSuTe1}. 
For early works on tachyon condensation in string theory,
see \cite{Halpern}.}
Therefore this unstable D0-brane picture gives unified view
for the all D-branes, including 
pure D-branes, the D-branes on the fuzzy sphere and 
the flat noncommutative D-branes.
Therefore, the system of the unstable D0-branes
gives a unified picture of the all D-branes
in the theory and could be a starting point to consider 
a nonperturbative definition of a string theory 
\cite{AsSuTe1}-\cite{AsSuTe3} \cite{TaTe} 
which can be regarded as a generalization of the 
matrix model \cite{BFSS, deWit}.\footnote{
The construction of the D$p$-brane from
the unstable D0-branes was shown only for the flat space-time
background although we expect generalizations 
to other background is possible.
In this paper, we will consider
the flat space-time
background only.}
Then, it is natural to ask 
how this unstable D0-brane picture 
incorporates the noncommutativity 
of the fuzzy D-branes, etc, 
which are represented as 
the mutually noncommutative matrix coordinates
in the BPS D0-brane picture.
This question is very interesting since 
the matrix coordinates 
of the unstable D0-branes corresponding 
to the D$p$-brane with the flux
are mutually commutative \cite{Te, AsSuTe3}
and there does not seem noncommutativity.

In this paper, 
we show that the tachyon condensation 
of the unstable D0-branes
induces the noncommutativity.
We will see that 
in the infinite tachyon condensation limit,
in which the corresponding D$p$-brane boundary state
becomes the usual form,
most of unstable D0-branes disappear by the tachyon condensation.
Indeed, the remaining D0-branes consist 
the BPS D0-branes with the correct noncommutative 
coordinates as we will see in the fuzzy $S^2$ case explicitly.
The noncommutativity appears
because of the disappearance of the D0-branes by
the condensation of the tachyon, which
does not commute with the matrix coordinates.

The fuzzy $S^2$ brane 
corresponding to the D2-brane on $S^2$ with a unit flux
should be built from one BPS D0-brane, however,
a BPS D0-brane can not be fuzzy.
Thus we might expect that there is no D2-brane picture 
for this ``fuzzy'' brane.
This might be a problem for 
the unified unstable D0-brane picture,
however, we also show that
the boundary state for the D$p$-brane on the curved manifold
which satisfies the Dirichlet boundary condition on the manifold
is not localized on the manifold.
Thus the D2-brane on $S^2$ with unit flux is
equivalent to a BPS D0-brane even though it is at the origin.
Here the D2-brane on $S^2$ means that 
it satisfies the Dirichlet boundary condition on the $S^2$,
but in reality it is at the origin.
Then we have another problem which is constructing
the D$p$-brane localized on the curved manifold.
Our result indicates that
the boundary state for 
the D$p$-brane localized on the curved manifold should have 
a nonzero boundary term which 
corresponds to 
the tachyon of the unstable D0-branes.
This implies that the D$p$-brane on the curved manifold
is ambiguous at the string scale.

The organization of the paper is as follows.
In section 2, we review 
how to construct D$p$-branes from 
infinitely many unstable D0-branes.
In section 3, 
we show for some cases the unstable D0-branes
are reduced to fewer number
of D0-branes by the tachyon condensation.
We consider the fuzzy $S^2$ case in which 
finite number of the D0-branes consist the D2-branes.
We also consider the flat noncommutative D$(2p)$-branes.
In section 4,
we show the tachyon condensation 
of the unstable D0-branes
induces the fuzziness to the world volume of 
the D$p$-brane.
We also discuss the diffeomorphism and Seiberg-Witten map
in terms of D0-branes.
Finally, we conclude our study in this paper and 
discuss some future problems in section \ref{concl}.
In appendix A, we briefly review the boundary states.
In Appendix B, we summarize the normalization of the 
boundary state. 
In Appendix C, we discuss the validity of the 
assumption made in section 3 and study
the Harmonic oscillator case.






\section{D$p$-brane from infinitely many unstable D0-branes}

In this section, 
we will explain how to construct
D$p$-branes from infinitely many unstable D0-branes.\footnote{
In this paper, we set $\alpha'=1$ or $\alpha'=2$
for bosonic string string or superstring theory, respectively,
unless we recover explicit $\alpha'$ dependence.}

First we will 
extend the construction of the Dp-branes from infinitely many
unstable D0-branes in superstring theory \cite{Te, AsSuTe3} 
to that in bosonic string theory.
The tachyons and other fields on $N$ D0-branes in bosonic string
become Hermitian $N \times N$ matrices. 
Taking a large $N$ limit, they may become
operators acting on some Hilbert space and 
each base element of an orthonormal basis of it correspond to
a D0-brane.
Let us consider the usual Hilbert space of quantum mechanics,
on which the operators $\wh p$ and $\wh x$ act.
Here they satisfy 
the usual commutation relation $[\wh x, \wh p]=i$.\footnote{
This system does not have normalized basis, however,
we can consider the harmonic oscillator system in Appendix \ref{hs}
as a regularization of the system. }
Note that this Hilbert space should be
distinguished from (the zero mode sector of) 
the closed string Hilbert space used in the boundary state.
Thus we will use two different types of operators
which act different Hilbert spaces.

Let us consider a following configuration of the tachyon $T$ and  
the massless scalars $\Phi^\mu$, 
which represent a position of the D0-branes:
\begin{equation}
 T=u^2 \wh H_0, \;\; \Phi^\alpha=\wh x^\alpha, 
\,\,(\alpha=1, \ldots,p),
\label{boscon}
\end{equation}
where $\wh H_0=\frac{1}{2} \sum_{\alpha=1}^{p} (\wh p_\alpha)^2$
and other fields were set to be zero.
Taking $u \rightarrow \infty$ limit,
the configuration (\ref{boscon})
will be localized near $\wh p_\alpha=0$
because the tachyon potential vanishes as $T \rightarrow \infty$
and the unstable D0-branes with $\wh p^2>0$ disappear. 
Actually the tachyon potential of the BSFT action \cite{KMM1, GeSh} is 
like $V(T)={\rm Tr} ( e^{-T}(1+T) )$
which goes to zero as $T \rightarrow \infty$.

This is a bosonic string analogue of 
the configuration of the unstable D$0$-branes in superstring theory
which represents the D$p$-brane \cite{Te}.
It can be obtained from 
the usual kink configuration 
$T=u x$ 
in a non BPS D$1$-brane
representing a BPS D0-brane
\cite{KMM2, KrLa, TaTeUe}
by replacing $x$ with $\wh p$ and 
taking $\Phi^\alpha=\wh x^\alpha$.
Indeed, in bosonic case, (\ref{boscon}) is also obtained
from the unstable kink configuration $T=u^2 x^2$ in 
D$25$-brane \cite{KMM1, GeSh}, which represents D$24$-brane, 
by the same procedure.
Therefore, in the same reason as in superstring case \cite{Te}, 
(\ref{boscon}) will be a D$p$-brane solution of the equations of motion 
for the BSFT action \cite{KMM1, GeSh} 
in the $u \rightarrow \infty$ limit.
In this paper, instead of using BSFT action,
we will use the boundary state to show this configuration
indeed represents the D$p$-brane
as in the superstring case \cite{AsSuTe1, AsSuTe2, AsSuTe3}.
See Appendix \ref{bsr} for a brief summary of 
the boundary state used in this paper.

The boundary state for the large $N$ D$0$-branes 
with (\ref{boscon}) is given by
\begin{eqnarray}
e^{-S_b} \ket{\! x \! = \! 0 \! }
= \TrP e^{-i\int d \sigma \hat{H}(\hat{x},\hat{p}) }\ket{\! x \! = \! 0 \! },
\end{eqnarray}
where
\begin{equation}
 i \wh H(\wh x,\wh p) =\frac{u^2}{2}
\sum_{\alpha=1}^{p} (\wh p_\alpha)^2
+i \wh x^\alpha P_\alpha(\sigma),
\end{equation}
$\ket{\! x \! = \! 0 \! }$ is the boundary state for a D0-brane at the origin
and $e^{-S_b}$ is the boundary interaction (\ref{D0bi}).
Since this is a quantum mechanical partition function
with Hamiltonian $\wh H$
which includes the time-dependent perturbation $i \wh x P(\sigma)$, 
we can rewrite it in terms of the path-integral formulation
as in \cite{AsSuTe3, Is}.
The Lagrangian corresponding to $\wh H$ is 
\begin{eqnarray}
 i L(x,\dot{x})=-\frac{1}{2 u^2} ({\dot{x}^\alpha(\sigma)})^2 
-i x^\alpha(\sigma) P_\alpha(\sigma),
\end{eqnarray}
and 
\begin{eqnarray}
e^{-S_b} \ket{\! x \! = \! 0 \! }
= \int [dx^\alpha(\sigma)] e^{i\int d \sigma L(x,\dot{x}) }
\ket{\! x \! = \! 0 \! }
=\int [dx^\alpha(\sigma)] e^{-\int d \sigma 
\frac{1}{2 u^2} (\dot{x}^\alpha)^2  }
\ket{x^\alpha(\sigma)}.
\label{hl}
\end{eqnarray}
Here, we set $p=25$ for simplicity, however, 
a generalization to a D$p$-brane is obvious.
Thus, in the $u \rightarrow \infty$ limit,
we have 
\begin{eqnarray}
\lim_{u \ra \infty} \, e^{-S_b} \ket{\! x \! = \! 0 \! } 
=
\int [dx^\alpha(\sigma)] \ket{x^\alpha(\sigma)}.
\end{eqnarray}
Because the boundary state for a D$25$-brane 
can be written as $\ket{D25}= \int [dx^\alpha(\sigma)] \ket{x^\alpha(\sigma)}$,
we have exactly proved that the D$p$-brane is 
equivalent to
the infinitely many D$0$-branes with (\ref{boscon}).

From the apparent similarity between this and
the one in \cite{AsSuTe3},  
we can easily incorporate the tachyon $t$, gauge field $A_\alpha$ 
and massless scalars $\phi^i$ ($i=p+1, \ldots, 25$)
on the D$p$-brane.
Actually, 
taking 
\begin{equation}
 T=u^2 \wh H_0 , \;\; \Phi^\alpha=\wh x^\alpha,
\;\; \Phi^i=\phi^i(\wh x),
\label{bosconf}
\end{equation}
with
\begin{equation}
 \wh H_0=\frac{1}{2} \sum_{\alpha=1}^{p} 
(\wh p_\alpha-i A_\alpha(\wh x) )^2
+t(\wh x),
\end{equation}
we have 
\begin{equation}
 i \wh H=\frac{u^2}{2} 
(\wh p_\alpha-i A_\alpha(\wh x) )^2 +t(\wh x)
+i \phi^i(\wh x) P_i   +i \wh x^\alpha P_\alpha,
\end{equation}
and a corresponding Lagrangian
\begin{eqnarray}
 i L(x,\dot{x})=-\frac{1}{2 u^2} \dot{x}^2 
-A_\alpha \dot{x}^\alpha -t( x)
-i \phi^i( x) P_i   -i x^\alpha P_\alpha. 
\end{eqnarray}
Thus, we obtain the correct boundary state 
of the D$p$-brane with $A_\alpha, t, \phi^i$:
\begin{eqnarray}
e^{-S_b} \ket{\! x \! = \! 0 \! } &=& 
\int [dx(\sigma)] e^{\int d \sigma \left( -\frac{1}{2 u^2} \dot{x}^2  
-A_\alpha \dot{x}^\alpha -t( x) \right) }
\ket{x^\alpha(\sigma), \phi^i(x^\alpha(\sigma))} \\
& \stackrel{u \ra \infty}{\longrightarrow}  & 
\int [dx(\sigma)] e^{\int d \sigma   
\left( - A_\alpha \dot{x}^\alpha -t( x) \right) }
\ket{x^\alpha(\sigma), \phi^i(x^\alpha(\sigma))}.
\end{eqnarray}
Therefore we have two different descriptions of 
the same physical system:
one using the infinitely many D$0$-branes
and the one using the D$p$-brane.
In other words, we can construct any D$p$-brane
from the D0-branes as in superstring case \cite{Te}.
Moreover, we can easily extend this construction of
a D$p$-brane to several D$p$-branes.
We can also generalize this construction
to curved D$p$-branes
by considering a Hamiltonian of a
particle on a curved space with gauge coupling and
a potential as in \cite{AsSuTe3}.
The fluctuations of the metric of the world volume 
can be included as in \cite{AsSuTe3}.
We will discuss this world volume metric and the diffeomorphism later.

Finally we will briefly summarize the result 
in \cite{AsSuTe1, AsSuTe3} for the superstring.
For simplicity, we consider $N$ $\dz$ pairs in type IIA 
although 
other cases, including non BPS D0-branes, were discussed 
in \cite{AsSuTe1, AsSuTe3}.
We always consider 
$N$ pairs of $\dz$
such that the $\Phi^\mu$ of a D0 and  
a $\overline{\mbox{D}0}$ in any pair of $\dz$ are same, i.e. 
the D0 and the $\overline{\mbox{D}0}$ in the pair
are at a same position.
The gauge symmetry for the $N$ pairs of $\dz$ is 
$U(N) \times U(N)$.
The action of the diagonal $U(N)$ part of it 
keeps the above condition 
for the position of $\dz$
and in what follows
$U(N)$ symmetry means this diagonal $U(N)$.

Let us consider the flat BPS D$(2p)$-brane \cite{Te}
in terms of $N$ $\dz$ pairs in the $N \ra \infty$ limit.
For this, the tachyon operator is given by $T=u D$ where
$D 
=\sum_{\alpha=1}^{2p} \gamma^\alpha 
(\wh p_\alpha-i A_\alpha(\wh x)))$,
which is the Dirac operator,
and $\Phi^\alpha=\wh x^\alpha$ $(\alpha=1, \ldots, 2p)$, 
$\Phi^i=0$ $(i=2p+1, \ldots, 9)$.
Then the boundary interaction  (\ref{D0D0bar})
is 
\beq
e^{-S_b}= 
\TrP \exp \int d \sigma \left( 
-u^2 (\wh p_\alpha-i A_\alpha(\wh x))^2 
+\f{u^2}{4} F_{\alpha \beta} [\gamma^\alpha, \gamma^\beta] 
-i \wh x^{\alpha} P_\alpha(\sigma) 
-i u \Pi_\alpha(\sigma) \gamma^\alpha 
\right),
\label{bif}
\eeq
which can be rewritten as
\beq
e^{-S_b}=
\int [dx^\alpha] [d \psi^\alpha]
\, {\rm P} \exp \int d\sigma
\left( -\f{\dot{x}_\alpha^2
+\psi^\alpha \dot{\psi}^\alpha}{4 u^2} 
-A_\alpha \dot{x}^\alpha 
+ \f{1}{2} F_{\alpha \beta} \psi^\alpha \psi^\beta
-i  x^{\alpha} P_\alpha 
-i \Pi_\alpha \psi^\alpha 
\right)
\label{subi}
\eeq
using the path-integral.
Thus, in the $u \ra \infty$ limit, 
we have the correct D$p$-brane boundary state
from the $\dz$ boundary state:
\beq
e^{-S_b} \ket{\! x \! = \! 0 \! } \ket{\psi=0}
 \ra 
\int [dx^\alpha] [d \psi^\alpha]
\, {\rm P} 
e^{ \int d\sigma
\left( 
-A_\alpha \dot{x}^\alpha 
+ \f{1}{2} F_{\alpha \beta} \psi^\alpha \psi^\beta
\right)}
\ket{x^\alpha,x^i=0} \ket{\psi^\alpha,\psi^i=0}.
\eeq

If we want to consider D$p$-brane 
on a curved submanifold in the flat space,
we should have the Dirac operator on it.
Using the spin connection 
\beq
\omega_{AB,\gamma}= e^{\alpha}_A e^{\beta}_B \left(
\Gamma^\delta_{\gamma \alpha} g_{\delta \beta}
-\delta_{CD} e^D_\beta \partial_\gamma e^C_\alpha   \right),
\eeq
where $e_{\alpha}^A$ is a vielbein satisfying
$e_\alpha^A \delta_{A B} e_\beta^B=g_{\alpha \beta}$,
the Dirac operator is defined by
\beq
D \equiv -i  \gamma^A e^\alpha_A \bigtriangledown_\alpha,
\label{diracop}
\eeq
where $\{ \gamma^A,\gamma^B \}=2 \delta^{A B}$
and
\beq
\bigtriangledown_\alpha=\f{\de}{\de x^\alpha} 
-\f{1}{8} \omega_{AB, \alpha} [ \gamma^A, \gamma^B ] 
-i A_\alpha.
\eeq
Then, we take
$T=u D$
and $\Phi^{\mu}=f^{\mu}(x^\alpha)$,
where $f^{\mu}$ is the embedding of the curved submanifold
to the flat space.
Inserting these into the formula of the boundary interaction 
(\ref{D0D0bar}),
we have 
\beq
e^{-S_b}= \TrP \exp \left( -i \int d \sigma  \wh H \right)
\eeq
with the total Hamiltonian
\beq
i \wh H = u^2 \wh H_0+ +i  f^{\mu}(x) P_\mu (\sigma)
+i u \Pi_\mu(\sigma) \f{\de f^\mu(x)}{\de x^\alpha} \gamma^\alpha,
\eeq
where 
\beq
\wh H_0= D^2.
\eeq
We can rewrite it in the path-integral formalism: 
\beq
e^{-S_b}=
\int [dx^\alpha] [d \psi^\alpha]
\, {\rm P} 
e^{ i\int d\sigma L },
\eeq
where
\beq
i L=-\f{1}{4 u^2} g_{\alpha \beta}(x) 
(\dot{x}^\alpha \dot{x}^\beta
+\psi^\alpha \bigtriangledown_\sigma {\psi}^\beta)
-A_\alpha \dot{x}^\alpha 
+ \f{1}{2} F_{\alpha \beta} \psi^\alpha \psi^\beta
-i  f^{\mu}(x) P_\mu 
-i \Pi_\mu \f{\de f^\mu(x)}{\de x^\alpha} \psi^\alpha,
\label{lag1}
\eeq
and $\bigtriangledown_\sigma {\psi}^\alpha
=\dot{\psi}^\alpha+\dot{x}^\beta 
\Gamma^\alpha_{\beta \gamma} \psi^\gamma$.
Thus in the $u \ra \infty$ limit, 
we obtain a boundary state 
for the D$p$-brane wrapping the submanifold:
\beq
\int [dx^\alpha] [d \psi^\alpha]
\, {\rm P} 
e^{ -\int d\sigma \left( A_\alpha \dot{x}^\alpha 
- \f{1}{2} F_{\alpha \beta} \psi^\alpha \psi^\beta \right) }
\ket{f^\mu(x)} \ket{\de_{\alpha} f^\mu \, \psi^\alpha }.
\label{b1}
\eeq
Here the path-integral measure $[dx^\alpha]$ may depend on 
the vielbein $e^A_\alpha$.
The measure may become usual one 
in a coordinate ${x'}^\mu(x)$
such that $g'_{\alpha \beta}(x') \sim \delta_{\alpha \beta}$.
Thus we expect that the boundary state (\ref{b1}) will be
the D$p$-brane boundary state 
if we choose the vielbein such that
the metric $g_{\alpha \beta}=e^A_\alpha e^A_\beta$
is the induced metric $\f{\de f^\mu}{\de x^\alpha}
\f{\de f^\mu} {\de x^\beta}$.

The normalization of the 
boundary states has not been considered. It will be discussed 
in Appendix \ref{normal}.

It should be emphasized that 
the boundary states we consider in this paper are
not BRST invariant in general.
Thus they do not represent solutions of the equations
of motion nor boundary conformal field theory in general
and not called boundary states in the strict sense of the word.
However,
these boundary states which are naively extended to off-shell field
are meaningful, at least, by considering 
the BSFT action for it which is obtained as
$S_{BSFT}=\f{2 \pi}{g_s} \bra{0} Dp \rangle$
and considering the couplings to the closed string fields,
such as the energy-momentum tensor.
In order to get the solutions, 
we should restrict the boundary integrations.
We will see some examples we consider in this paper
are on-shell and the boundary states are BRST invariant
although we do not consider the restriction in general cases.

\section{Disappearance of D0-branes and Fuzzy D-branes
}
\label{red}

In this section, 
we will consider a Hamiltonian $\wh H_0$
which has a gap above ground states.\footnote{
The spectrum of the free Hamiltonian is gapless.
In this case,
we can deform the Hamiltonian 
such that the Hamiltonian has a gap and
then consider
the free Hamiltonian as a limit of the deformed one.
Indeed, we will consider a
harmonic oscillator Hamiltonian, which can be regarded as 
a IR regularization of the free Hamiltonian
in the Appendix \ref{hs}.}
We take an orthonormalized eigen state $\ket{n}$ of $\wh H_0$ 
with the eigen value $E_n$, i.e.
$\wh H_0 \ket{n}=E_n \ket{n}$ and $\bra{n} n' \rangle =\delta_{n, n'}$.
The eigen states span the
whole Hilbert space and each eigen state represents
an unstable D0-brane.
Note that 
we can choose any base of the Hilbert space.
The orthonormalized bases are related each other
by a $U(N)$ gauge transformation of the unstable $N$ D0-branes.
Especially, the path-integral formalism 
is closely related to a position basis $\{ \ket{x^\mu} \}$. 
At $u=0$ we can take $\{ \ket{x^\mu} \}$ basis
and this system is a collection of $N$ D0-branes
which have a definite position.

In the $u \rightarrow \infty$ limit, 
we expect that only the D$0$-branes which correspond to 
the ground states of $\wh H_0$ contribute 
in the boundary interaction
\begin{eqnarray}
e^{-S_b}=\TrP e^{-i\int d \sigma \hat{H}(\hat{x},\hat{p}) }
\label{bp1}
\end{eqnarray}
because other D0-branes have infinitely large tachyon values
and disappear by the tachyon condensation.\footnote{
For the bosonic case
we have chosen the constant part of $t$ such that 
the eigen value of $T$ of ground states is zero or positive.
Note that $t$ possibly depends on $u$.}
Denoting the degeneracy of the ground state
as $N_0$ and the ground states as $\ket{a}$, $(a=1,\ldots, N_0)$,
this means 
we will have 
\begin{eqnarray}
e^{-S_b}=\Tr_{N_0 \times N_0} {\rm P} e^{-i\int d \sigma 
\tilde{\Phi}^\mu P_\mu(\sigma)
}
\label{bp2}
\end{eqnarray}
in bosonic string or
\begin{eqnarray}
e^{-S_b}=\Tr_{N_0 \times N_0} {\rm P} e^{-i\int d \sigma \,
\left( 
\tilde{\Phi}^\mu P_\mu(\sigma)
+\f{1}{2} [\tilde \Phi^\mu, \tilde \Phi^\nu] 
\Pi_\mu(\sigma)\Pi_\nu(\sigma)
 \right) }
\label{bp2a}
\end{eqnarray}
in superstring,
where an $N_0 \times N_0$ matrix $\tilde{\Phi}^\mu$ is 
given by 
\beq
(\tilde{\Phi}^\mu)_{ab}= \bra{a} \Phi^\mu(x^\alpha) \ket{b},
\eeq
which represents the $N_0$ D0-branes 
with matrix coordinates $\tilde{\Phi}^\mu$.\footnote{
In (\ref{bp2}) we have set $\phi^i(x)=0$.
If we include this, 
the exponent of the left hand side of (\ref{bp2}) 
will become 
$\tilde{\Phi}^\mu P_\mu(\sigma)+\tilde{\phi}^i P_i(\sigma)$
where $(\tilde{\phi}^i)_{ab}=\bra{a} \phi^i(\wh x^\alpha) \ket{b}$.}

In this paper, we assume this reduction is indeed true.
(In Appendix \ref{hs} we will discuss
the validity of this assumption for some cases.)
Then, 
the the original $N$ unstable D0-branes with the boundary interaction
is equivalent to the $N_0$ D0-branes with 
$(\tilde \Phi^\mu)_{ab}=\bra{a} \Phi^\mu \ket{b}$
$(a,b=1,\ldots, N_0)$
in this limit.
Note that 
for the $\dz$ case 
we likely have 
only $N_0$ $D0$ or only $N_0$ $\overline{\mbox{D}0}$
after the tachyon condensation
and then there is no tachyon on $N_0$ D0-branes.

As a simplest example of this phenomenon, let us 
consider two D0-branes in bosonic string theory
and take 
\beq
T=u^2 ({\rm 1}_{2 \times2}+\sigma_1), \,\,
\Phi^1=a \, \sigma_3
\label{twoD0}
\eeq
(or 
two non BPS D-branes with 
$T=\f{u}{\s{2}} ({\rm 1}_{2 \times2} +\sigma_1)$
and $\Phi^1=a \sigma_3$
in the type IIB superstring).
At $u=0$, two D0-branes are at 
$x^1=a$ and $x^1=-a$.
After the unitary transformation $U=\f{1}{\s{2}} (1+i \sigma_2)$, 
we have a diagonal 
tachyon $T=u^2 \, {\rm diag}(2,0)$
and $\Phi^1=a \, \sigma_1$.
For $0 < u^2 < \infty$, the system becomes 
fuzzy D-branes extending around $-a<x^1<a$.
Then at $u=\infty$, only a D0-brane 
located at $x^1=0$ remain
since the D0-brane corresponding to 
the nonzero eigen state of $T$
disappears by the tachyon condensation
and $\bra{0} \Phi^1 \ket{0}=0$ where
$\bra{0}=(0,1)$ is the ground state.
We can easily see that 
only the fluctuations proportional to $\ket{0}\bra{0}$ 
correctly survive at $u =\infty$.
This example clearly shows
that the D-brane becomes fuzzy 
even with commutative $\Phi^\mu$
if the tachyon is turned on.
Note that $a$ can be taken very large 
so that they are separated far away each other
at $u=0$.
We also note that
this is the solution of the equqtions of motion of D0-branes
only for $u=0$ or $u=\infty$.
We will discuss this noncommutativity deeper
in section \ref{nct}.

It is more interesting to 
start from infinitely many unstable D0-branes.
We have seen that If $\wh H_0$ is a Hamiltonian
for a particle on a manifold,
there is another description in terms of 
a D$p$-brane.
Therefore we conclude that 
in the $u \ra \infty $ limit
the $N_0$ D0-branes with 
$(\tilde \Phi^\mu)_{ab}=\bra{a} \Phi^\mu \ket{b}$
$(a,b=1,\ldots, N_0)$
are equivalent to
the D$p$-brane.
In particular, 
the Hamiltonian corresponding to
the D$p$-brane with flux on its world volume
typically has a gap between the ground states,
namely, lowest Landau level,
and excited states.
For examples, the flat D$(2p)$-brane with flux
is the noncommutative D$(2p)$-brane which
has a description in terms of 
infinitely many D0-branes with noncommutative coordinates
and
the D2-brane on $S^2$ with $n$ flux
is the fuzzy D2-brane.

The relation between 
the D$p$-brane with flux and the D0-branes
were studied recently in \cite{El}.
The boundary states used in \cite{El} for superstring case
look very similar to those in \cite{AsSuTe1, AsSuTe3}.
In fact, we can see that
they are same as the boundary states in \cite{AsSuTe1, AsSuTe3}
although 
the $u$ dependent term in $\wh H$ 
was regarded as an auxiliary regularization term
in \cite{El}, in which 
the unstable D0-branes with tachyons
were not considered.
Thus, we can easily translate the results
in \cite{El} into our unstable D0-branes picture
with some changes of notations.
Especially, 
the D2-brane on $S^2$ with $n$ flux in bosonic string theory
is an interesting example
since the $N_0=|n|+1$ is finite.

In the following, we will consider
a D2-brane on $S^2$ with $n$ flux 
in the flat space-time 
in terms of infinitely many $\dz$ branes
in type IIA superstring theory
as an example of the equivalence.
This provides an interesting example
for considering the fuzziness
induced by the tachyon condensation 
since the D2-brane with a unit flux 
should be equivalent to one BPS D0-brane
which obviously is not on $S^2$, but at the origin.

\subsection{Fuzzy $S^2$ Brane from Tachyon Condensation}
\label{s2}

First, let us construct
the Dirac operator of charged spin $\f{1}{2}$ particle in
a constant magnetic field on $S^2$ 
(See, for example, \cite{Re, Ab}).
Taking the coordinate $(\theta, \phi)$ as usual,
the dreibein is $e^i_\mu={\rm diag} (1, \sin \th)$
and the gauge field in the singular gauge 
is given by $A_\phi=\f{n}{2} \cos \theta$
where $n$ should be an integer.
Thus the Dirac operator (\ref{diracop}) in this case is 
\beq
D=-i \sigma_x 
\left( \de_\theta+\f{1}{2} \f{\cos \theta}{\sin \theta} \right)
-i \sigma_y  \f{1}{\sin \theta} 
\left( \de_\phi - i \f{n}{2} \cos \theta \right)
\eeq
where $\sigma_x,\sigma_y$ are the Pauli matrices.
We can compute the $D^2$, which can be regarded as the Hamiltonian
of the supersymmetric quantum mechanics, as
\beq
D^2=-\f{1}{\sin \th} \de_\th \sin \th \de_\th
-\f{1}{\sin^2 \th} (\de_\phi-i \f{n}{2} \cos \th)^2
+i \sigma_z \f{\cos \th}{\sin^2 \th} 
(\de_\phi - i \f{n}{2} \cos \th)
+\f{1}{4}(1+\f{1}{\sin^2 \th}) +  \sigma_z \f{n}{2}.
\eeq

This system has the three-dimensional rotation symmetry 
which allow us to define the angular-momentum operators;
\beqa
L_+ &=& e^{i \phi} \left( \f{\de}{\de_\th} +i \f{\cos \th}{\sin \th} 
\f{\de}{\de_\phi} +\f{1}{2 \sin \th} (\sigma_z +n) \right),   \\ 
L_- &=&  e^{-i \phi} \left( -\f{\de}{\de_\th} 
+i \f{\cos \th}{\sin \th} 
\f{\de}{\de_\phi} +\f{1}{2 \sin \th} (\sigma_z +n) \right),  \;\;
L_z=-i \f{\de}{\de_\phi}, 
\eeqa
which satisfy $[ L_+, L-]=2 L_z, [ L_z, L_\pm ]=\pm L_\pm$
and $0=[D,L_z]=[D,L_\pm]$.
As in \cite{Haldane},
we can rewrite the Hamiltonian as
\beq
D^2=L^2+\f{1}{4}-\f{n^2}{4}.
\eeq
By diagonalizing $L^2$ and denoting $L^2=j(j+1)$ as usual,
we obtain 
\beq
D^2=\left(j+\f{1}{2}\right)^2-\f{n^2}{4},
\eeq
which means $j+\f{1}{2} \geq |\f{n}{2}|$
because $D$ is a hermitian operator.
Therefore, the ground states are expected to 
have $j=|\f{n}{2}|-\f{1}{2}$
and the degeneracy of the ground state is $2j+1=|n|$.

It is easy to find the zero-mode of $D$.
Since the $D \sigma_x$ is a diagonal matrix,
$D\psi(\theta,\phi)=0$ is equivalent to
\beq
\left( \de_\th + \f{1}{2} \f{\cos \th}{\sin \th}
\pm \f{m-\f{n}{2} \cos \th}{\sin \th}   \right) \psi_\pm(\th)=0,
\eeq
where 
$\psi(\theta,\phi)=\, {}^t (\psi_-(\th) e^{im\phi}, 
\psi_+ (\th) e^{im\phi} )$.
Thus we obtain
\beq
\psi_\pm(\th)= {\rm const. } \,\,\,
(1-\cos \th)^{\f{1}{4}(-1\mp 2m \pm n)} 
(1+\cos \th)^{\f{1}{4}(-1\pm 2m \pm n)},
\eeq
which is nonsingular if and only if
$-1\mp 2m \pm n \geq 0$ and 
$-1\pm 2m \pm n \geq 0$.\footnote{
Note that the wave function  $\psi(\theta,\phi)$ is singular 
if $-1\mp 2m \pm n = 0$ or $-1\pm 2m \pm n = 0$
at $\theta=0$ or $\theta=\pi$, respectively.
However, these singularities can be removed by regular 
gauge transformations, therefore these are harmless.}
This implies $\pm n  \geq 1$ and then we should set
$\psi_-=0$ or $\psi_+=0$ for $n>0$ or $n<0$, respectively.
Note that there is no zero mode of 
the Dirac operator for $n=0$.
This is, of course, consistent with the Lichnerowicz formula.
For given $n$, we can take 
\beq
|n|-1 \geq 2m \geq -(|n|-1),
\eeq
and $2m$ should be even or odd integer if $|n|-1$ 
is even or odd integer, respectively, 
since we are considering the spin $\f{1}{2}$ particle.
Therefore we correctly obtain
$|n|$ Dirac zero modes, which we will denote $\ket{m}$,
$(m=-(|n|-1)/2, -(|n|-3)/2, \cdots, (|n|-3)/2, (|n|-1)/2)$.

Let us consider the $\dz$ pairs in type IIA string theory and
take $T= \f{u}{R} D$ and $\Phi^1=R \sin \th \cos \phi, \, 
\Phi^2=R \sin \th \cos \phi, \, \Phi^3=R \cos \th$.
In the limit $u \rightarrow \infty$,
only the Dirac zero modes $\ket{m}$ survive
and the $\dz$ pairs corresponding to
massive modes of the Dirac operator disappear 
by the tachyon condensation.
Therefore we have $|n|$ D0-branes or 
$|n|$ $\overline{\mbox{D}0}$-branes for $n>0$ or $n<0$,
respectively.
(Here we think $-\sigma_3$ as the bilinear 
of the boundary fermions 
$[ \eta, \bar{\eta}]$ in \cite{TaTeUe}.)
The positions of 
these D0-branes, 
$|n| \times |n|$ matrices $(\tilde \Phi^\alpha)_{m m'}$,
are given by
\beqa
(\tilde \Phi^1)_{m m'} &=&  R \bra{m} \sin \th \cos \phi \ket{m'}, \CR
(\tilde \Phi^2)_{m m'} &=&  R \bra{m} \sin \th \sin \phi \ket{m'}, \CR
(\tilde \Phi^3)_{m m'} &=&  R \bra{m} \cos \th \ket{m'}.
\eeqa
Applying the Wigner-Eckart theorem,
we know that $\tilde \Phi^\alpha_{m m'}$ is proportional to
the generators of the $|n|$ dimensional representation
of $SU(2)$, $(J^\alpha)_{m m'}$.
By comparing $\bra{m_{max}} \cos \th \ket{m_{max}}$ and 
$\bra{m_{max}} L_z \ket{m_{max}}$, 
we obtain
\beq
(\tilde \Phi^\alpha)_{m m'}={\rm sign}(n) \,  
\f{2 R}{|n|+1} (J^\alpha)_{m m'}.
\label{loc}
\eeq
Note that
\beq
( (\tilde \Phi^1)^2+(\tilde \Phi^2)^2+(\tilde \Phi^3)^2 )_{m m'}
=R^2 \f{|n|-1}{|n|+1} \delta_{m m'}
\label{radius1}
\eeq
and $[\tilde \Phi^1, \tilde \Phi^2]=
{\rm sign}(n) \f{4 i R}{|n|+1} \tilde \Phi^3$.
It is interesting to see the effective radius of the $S^2$
defined from (\ref{radius1}), 
$R_{eff} \equiv R \sqrt{\f{|n|-1}{|n|+1}}$, 
is always smaller than $R$.

On the other hand, 
according to the previous discussion,
we can use the path-integral formalism
to represent this $\dz$ system and
obtain a BPS D2-brane on 
the two-sphere of radius $R$ 
with the uniform $n$ flux.
Because this D2-brane description and 
the D0-brane description of the same system 
should be equivalent,
we conclude that
the 
BPS D2-brane on the two-sphere 
with the uniform $n$ flux is equivalent
to the $n$ BPS D0-branes
with (\ref{loc}) \cite{KaTa} exactly.
We note that
this equivalence holds for any order in $\alpha'$,
i.e. beyond the approximation used in the DBI action,
and
the number of BPS D0-branes is correct even for finite $n$
while we have $|n|+1$ ground states 
for $n$ flux 
in the bosonic string \cite{El}. 
Of course, this is a consequence of the D0-brane charge conservation
and is related to the Atiyah-Singer index theory 
as shown in \cite{AsSuTe3}. 

For $n=1$, we have only a BPS D0-brane 
with $\Phi^\alpha=0$, which is localized at the origin.
However, in the D2-brane picture,
the D-brane seem to be localized on the $S^2$ which
does not contain the origin.
Moreover, for $n=0$, there are nothing, i.e.
D-branes completely disappear by the tachyon condensation.
These seem strange in the D2-brane picture, however,
we will see these are consistent
in section \ref{nct}.

It is important to note that
this fuzzy $S^2$ brane is off-shell
except for $|n|=1$.
In order to get the solution of the equations of motions,
we have to turn on the RR-flux as in \cite{Myers}.
This might not change the above discussion essentially
since the RR-flux considered in \cite{Myers}
only change the Chern-Simons term.
If the RR flux is small, $R$ will be proportional to
the strength of the RR flux.
See also \cite{HiNoSu}.
If we want to consider the on-shell fields,
we can consider the D3-D1 bound state where 
the fuzzy $S^2$ appear in the D1-brane world volume theory
near the D3-branes.
In this case
it may be the solution of the equations of motions
and the problem discussed here still appears,
i.e. pictures in the D3-brane and the D1-branes seem different,
especially for $|n|=1$.
It is also resolved in the same way discussed in this paper \cite{HaTe3}.

\subsection{Flat Noncommutative D-brane}

In this subsection, we will consider
the flat noncommutative D$(2p)$-brane.
In \cite{Is}, 
it was shown that
the boundary state of
the flat D$(2p)$-brane with the constant
field strength $F$ on the world volume
is equivalent to
the boundary state of 
the infinitely many D0-branes with 
$[\Phi^\alpha, \Phi^{\beta}]=i \left(\f{1}{F}\right)^{\alpha \beta}$.
This can be easily extended to the supersting case \cite{AsSuTe3}.

On the other hand, 
as we reviewed in section 2 we know that
the flat D$(2p)$-brane with the constant
field strength $F$ is also equivalent to
the infinitely many unstable D0-branes
with the non-trivial tachyon condensation \cite{AsSuTe3}.
Combining these two equivalences,
we can conclude 
the infinitely many unstable D0-branes
with the non-trivial tachyon condensation
are equivalent to the infinitely many BPS D0-branes with 
$[\Phi^\alpha, \Phi^{\beta}]=i \left(\f{1}{F}\right)^{\alpha \beta}$.
Now we can directly derive this equivalence
by considering the disappearance of the unstable D0-branes
by the tachyon condensation.
Indeed, as noted before, 
this was essentially done in \cite{El}.
For example in type IIA string theory,
the tachyon of the unstable $\dz$ branes is
the Dirac operator for the flat space with
constant field strength which 
has a gap between the lowest Landau level and excited states.
Actually, the Hamiltonian is 
\beq
\wh H_0={\rm 1}_{2 \times2} ( (\wh p_1-F \wh x_2)^2+\wh p_2^2 ) 
-\sigma_3 F_{12},
\eeq
and
the lowest Landau level for $F_{12}>0$ is 
\beq
\ket{k} \sim  \left( \begin{array}{c} 1 \\ 0 \end{array} \right)
\exp \left( ik x^1-\frac{F}{2} (x^2-k/F)^2 \right)
\eeq
in a gauge condition.
Therefore in the $u \ra \infty$ limit, 
we have infinitely many BPS D0-branes 
(or $\overline{\mbox{D}0}$-branes) which corresponds
to the lowest Landau level out of 
the infinitely many $\dz$ pairs.
The states in the lowest Landau level are parametrized
by $p$ real numbers which correspond to momenta,
and, as shown in \cite{El},
the coordinates of the remaining D0-branes
become
\beq
(\tilde{\Phi}^1)_{k k'}=-i \f{\de}{\de k} \delta(k-k'),
\;\;\; (\tilde{\Phi}^2)_{k k'}=\f{k}{F} \delta(k-k'),
\eeq
where we considered $p=1$ case for simplicity.
This correctly reproduce
the commutation relation 
\beq
[ \tilde{\Phi}^1, \tilde{\Phi}^2 ]= i \f{1}{F_{12}}.
\label{comm1}
\eeq
Note that 
it becomes the solution of the equation of motion
in the $u \rightarrow \infty$ limit
since we know the configuration (\ref{comm1}) is on-shell.

The flat noncommutative D-brane is also obtained from
the fuzzy D-brane on $S^2$ by taking some limit.
In fact,
taking $n \ra \infty$ and $\gamma \equiv \f{4 R^2}{n}$ fixed,
we have $[ \tilde{\Phi}^1, \tilde{\Phi}^2 ]= i \gamma$
near $\tilde{\Phi}^3 \sim R$, $\tilde \Phi^1 \sim 
\tilde \Phi^2 \sim 0$.

\section{Noncommutativity and Tachyon of D0-branes}
\label{nct}

In this section, we will see the tachyon condensation 
of the unstable D0-branes
induces the fuzziness to the world volume of 
the D$p$-brane where
the parameter $u$ controls 
the fuzziness.
We will also see that
the D$p$-brane is commutative and localized for $u \ra 0$
and becomes noncommutative for $u >0$.

Let us consider the $S^2$ case \cite{KaTa, Myers},
which we have studied in 
the the subsection \ref{s2}, for illustrative purposes since
generalizations to other cases are straightforward.

At $u=0$, there is no tachyon condensation on 
the unstable D0-branes.
Therefore
the boundary state is a simple sum
of the boundary states of the D0-branes which
have definite positions and 
each D0-brane corresponds to 
an $\wh x^\mu$ eigenstate.\footnote{
At $u=0$, the energy density is 
$C N/(g_s V)$, where $C$ is a numerical factor,
$N$ is 
the number of the unstable D0-branes and $V$
is the volume of the D$p$-brane constructed from D0-branes.
Thus it is infinite at $u=0$.
In order to avoid this difficulty,
we can 
take $u$ very small, but finite 
or approximate the Hilbert space as 
a finite dimensional space.
Actually, we can take the Hilbert space spanned
by the energy eigen states such that their energies are 
lower than some fixed energy. }
These D0-branes are commutative and
are localized on 
the sphere $\sum_{\mu=1}^3 (x^\mu)^2=R^2$.

Once we turn on nonzero $u$,
the exponent
in the boundary interaction 
contains the operators which are noncommutative each other, namely
the tachyon does not commute
with $\Phi^\alpha=\wh x^\alpha$.
This noncommutativity between 
the tachyon and the position operators
leads the noncommutativity or the fuzziness
of the D$p$-brane world volume.
Actually, 
replacing $P^\mu(\sigma)$ to its zero mode 
$\hat{p}_0^\mu$,
we find that
\beq
\left( 
\sum_{\mu=1}^3 (\hat x_0^\mu)^2-R^2
\right) 
e^{-S_b}|_{P(\sigma)=\hat p_0} \ket{\! x \! = \! 0 \! }
=\left( 
\sum_{\mu=1}^3 (\hat x_0^\mu)^2-R^2
\right) 
\Tr \left( e^{-i \Phi^\mu \hat p_0^\mu
-2 \pi u^2 D^2} \right) \ket{\! x \! = \! 0 \! } \neq 0.
\eeq

For $u \gg 1$, 
the eigen states of $D$ 
are more appropriate basis than 
the eigen states of $\wh x$.
For $u=\infty$,
we have $n$ BPS D0-branes 
which have non-commutative coordinates $\tilde \Phi^\mu$
(for $n>1$) as we have seen in section 3.
We can compute 
the charge density, the stress-energy tensor and other sources 
for the closed string modes which are expressed 
as $\bra{0} V e^{-S_b} \ket{\! x \! = \! 0 \! }$
in the boundary state formalism
where $V$ is 
the polynomials of the creation operators corresponding to the 
closed string modes \cite{roll}.
Indeed,
it was shown in \cite{Ha} that the D0-brane charge density
of the $n$ BPS D0-branes 
is not localized on $S^2$ of radius $R$ for finite $n$
using the formula 
of the matrix model
\cite{TaRa}.
(See also \cite{Sa}.)
Note that we can not have the D0 charge density localized on $S^2$ of radius $R$ 
even if we include 
the possible corrections to the charge density
discussed in \cite{Ha}.
This is obvious for $n=1$ case because all the corrections 
are commutators and vanish for $n=1$.

In the path-integral formalism,
we use c-numbers instead of the operators.
This seems to imply that
the D$p$-brane is commutative and localized
on the $S^2$ irrespective of $u$.
However, this is not true for $u>0$.
The reason is as follows.
For the $S^2$ case,
the path-integral variables are $\theta(\sigma)$
and $\phi(\sigma)$.\footnote{
Here we ignore the fermionic variables
since they are irrelevant for the discussion in this section.}
The coupling 
$i x^\alpha(\theta(\sigma),\phi(\sigma)) P_\alpha(\sigma)$  
in the Lagrangian $L(x,\dot{x})$
gives a factor 
\beq
\exp\left(-i 
\left( \int d\sigma x^\alpha(\theta(\sigma),\phi(\sigma)) \right)
\hat{p}_0^\alpha \right),
\label{factor1}
\eeq
to the boundary state.
Here we are concentrating on the zero mode $\hat{p}_0$
of $P(\sigma)$ and dropping the non-zero modes 
in order to see the space-time distribution
of the D$p$-brane.

In the $u \ra 0$ limit 
we can ignore the non-zero modes
and D$p$-brane is localized.
This is because 
the $u$-dependent term 
$-\f{1}{2 u^2} g_{\alpha \beta} \dot{x}^\alpha \dot{x}^\beta$
in the Lagrangian becomes infinite in the limit
and then the non-zero modes of 
$\theta(\sigma)$ and $\phi(\sigma)$ 
becomes infinitely ``massive'' and are decoupled.
Then we will have
$\exp\left(i 
\left( \int d\sigma x^\alpha(\theta(\sigma),\phi(\sigma)) \right)
P^\alpha(\sigma) \right)
=\exp\left(i x^\alpha(\theta_0,\phi_0) \hat{p}_0^\alpha \right)$,
where $\theta_0,\phi_0$ are the zero modes of 
$\theta(\sigma),\phi(\sigma)$,
therefore,
the D$p$-brane is localized on the $S^2$.
This is consistent with
the result in the D0-brane picture.

However, for $u>0$ 
the non-zero modes of $\theta(\sigma)$ and $\phi(\sigma)$ 
survive in 
$\int d\sigma x^\alpha(\theta(\sigma),\phi(\sigma))$
because $x^\alpha$ is not linear in 
$\theta(\sigma)$ and $\phi(\sigma)$. 
Thus the integrations of the non-zero modes of
$\theta(\sigma)$ and $\phi(\sigma)$ make
the $\hat{p}_0$ dependent factor 
(\ref{factor1})
very complicated,
which may give the noncommutativity or fuzziness
of the D$p$-brane world volume!
Note that the non-zero modes of $P(\sigma)$
are coupled to $x(\sigma)$ and then 
the space-time distribution of the D$p$-brane depends on 
what we probe it with. 
(For example, we will have different space-time distributions
using the couplings to dilaton, graviton, RR fields, and so on.)

This consideration can be clearly applied to any 
curved D$p$-brane.
On the other hand, for the flat D$p$-brane, 
$x^\alpha(\sigma)$ itself
is the path-integration
variable and $\int d \sigma x^\alpha(\sigma)=x^\alpha_0$. 
Therefore the boundary state has
the factor
$\exp\left(i x^\alpha_0 \hat{p}_0^\alpha \right)$
which mean D$p$-brane is localized on 
the hyperplane for any $u$ as expected.
Nevertheless, we can still think
it is noncommutative for $u>0$.
Actually, in the D0-brane picture,
the constituents D0-branes becomes 
noncommutative when $u>0$ even for the flat case.
The point is that
the flat D$p$-brane looks commutative 
since the noncommutativity is only within the hyperplane.
In the BPS limit $u \ra \infty$,
the noncommutative parameter
becomes $1/F$ when we turn 
the flux on the D$p$-brane world volume.
This noncommutativity becomes infinite 
in the $F \ra 0$ limit,
therefore the flat D$p$-brane can be considered to have 
a maximal noncommutativity in the D0-brane picture.
Note that 
we can think the world volume of 
the D$p$-brane with the flux 
is either commutative or noncommutative
because of the Seiberg-Witten map.
The picture here is consistent with 
the latter one.

Let us study the boundary conditions which
the boundary state satisfies.
Using $X^\mu(\sigma) \ket{\! x \! = \! 0 \! }=0$ and
the $X^\mu(\sigma)=i\f{\de}{\de P^\mu(\sigma)}$,
we find
\beq
\lim_{\epsilon \ra 0} 
\left( \left( \sum_{\mu=1}^3 X^\mu(\sigma+\epsilon) X^\mu(\sigma)
-R^2 \right)
e^{-S_b} \ket{\! x \! = \! 0 \! } \right)
=0,
\label{dc}
\eeq
for any finite $u$, which means that 
the boundary state satisfies 
the Dirichlet condition on the $S^2$ with the radius $R$.
In the path-integral formalism 
(\ref{dc}) is obvious.
In the operator formalism,
we also see that
(\ref{dc}) is satisfied
because of the path-ordering in the trace.

On the other hand, in the $u \ra \infty$ limit
we have (\ref{radius1}) which means
that the effective radius of $S^2$ which 
D2-brane wraps is less than $R$.
This seems puzzling, however,
(\ref{dc}) does not mean
$
\left( \sum_{\mu=1}^3 (X^\mu_0)^2
-R^2 \right)
e^{-S_b} \ket{\! x \! = \! 0 \! }
=0,
$
where $X^\mu_0=\f{1}{2\pi}\int d \sigma X^\mu(\sigma)$,
except for $u \ra 0$,
so the D2-brane does not located 
on the $S^2$ with the radius $R$ in spite of (\ref{dc}).
Actually, 
the average $\bar{x}^\mu
=\f{1}{2\pi} \int d\sigma x^\mu(\sigma)$ 
of $x^\mu(\sigma)$ satisfying
$\sum_{\mu=1}^3 (x^\mu(\sigma))^2=R^2$ always 
satisfies $\sum_{\mu=1}^3 (\bar x^\mu)^2<R^2$
except it is a constant map,
so we can intuitively understand 
why the effective radius of $S^2$ is less than $R$.

There is another problem for $|n|=1$.
In this case, we have 
$
X^\mu(\sigma) e^{-S_b} \ket{\! x \! = \! 0 \! }=0 , \,\, (\mu=1, \ldots, 3)
$
in the $u \ra \infty$ limit
since the boundary state is a BPS D0-brane at the origin.
This seems to contradict (\ref{dc}).
However, (\ref{dc}) is not valid in the $u \ra \infty$ limit,
namely, the $\ep \ra 0$ limit and the $u \ra \infty$ limit
do not commute each other.
Therefore there is no contradiction.
In order to see the two limits do not commute 
each other explicitly, 
let us consider 
the two unstable D0-branes system (\ref{twoD0}).
The boundary state for this satisfies
\beq
\lim_{\epsilon \ra 0} (X(\sigma+\epsilon) X(\sigma) -1)
\, e^{-S_b}\ket{\! x \! = \! 0 \! }=0
\eeq
for finite  $u$ 
and $X(\sigma) \, e^{-S_b}\ket{\! x \! = \! 0 \! }=0$ for $u \ra \infty$.
At the lowest order in $P(\sigma)$,
we have
\beq
X(\sigma+\epsilon) X(\sigma) e^{-S_b} \ket{\! x \! = \! 0 \! }=
\Tr \left(
e^{-u^2  (2 \pi -\epsilon) {\rm diag} (2,0)}
\sigma_1 e^{-\epsilon u^2 {\rm diag} (2,0)}
\sigma_1
\right) \ket{\! x \! = \! 0 \! }+{\cal O}(P(\sigma)).
\eeq
Taking $u \ra \infty$ with keeping 
$\beta \equiv \epsilon u^2$ finite,
we have 
\beq
X(\sigma+\epsilon) X(\sigma) e^{-S_b} \ket{\! x \! = \! 0 \! }
=e^{-2 \beta} \ket{\! x \! = \! 0 \! }
+{\cal O}(P(\sigma)).
\eeq
Thus 
for very large, but finite $u$,
$\lim_{\epsilon \ra 0} (X(\sigma+\epsilon) X(\sigma) -1)
\, e^{-S_b}\ket{\! x \! = \! 0 \! }  \sim 0+{\cal O}(P(\sigma))$.
However, if we take $u \ra \infty$ first, which
means $\beta \ra \infty$,
$\lim_{\epsilon \ra 0} (X(\sigma+\epsilon) X(\sigma) )
\, e^{-S_b}\ket{\! x \! = \! 0 \! }  = 0+{\cal O}(P(\sigma))$
which is consistent with $X(\sigma) \, e^{-S_b}\ket{\! x \! = \! 0 \! }=0$.
From this observation, 
we expect
for $\epsilon \ll 1/u^2 \ll 1$
we have $(X(\sigma+\epsilon) X(\sigma) -1)
\, e^{-S_b}\ket{\! x \! = \! 0 \! }  \sim 0$ and
for $1 \gg \epsilon \gg 1/u^2 $
we have $X(\sigma+\epsilon) X(\sigma) 
\, e^{-S_b}\ket{ \! x \! = \! 0  \!}  \sim 0$. 
Note that the $X(\sigma)$ is not a function of $\sigma$,
but an operator valued function of $\sigma$.


In summary, 
we have a (almost) commutative D$p$-brane 
for $u \ll 1$.
As $u$ goes larger the fuzziness of the world volume
becomes stronger and
the D0-brane picture is more natural for $u \gg 1$.
An interesting conclusion is that
a curved D$p$-brane which 
has Dirichlet boundary condition 
on a curved submanifold is not localized on it.
This fact resolves the problem for the D2-brane
on $S^2$ with the unit flux.


Some comments are as follows.
Taking $F \ra \infty$ limit
for the flat noncommutative D-brane
or $n \ra \infty$ limit with $R$ fixed for the fuzzy $S^2$ D-brane,
we have 
a similar situation as in $u \ra 0$ limit,
i.e. the localized D-brane.
Actually, using the D$p$-brane picture,
we can see that these have infinite energy because of the 
infinite field strength, which means that
the BPS D0-branes are dominated and 
other D0-branes which disappear by the tachyon condensation
are not important.
In fact, the energy gap between the ground states and
excited state is roughly proportional to $F$ or $n$.
Thus, for finite $u$, the value of $u$ 
is not important for $F \ra \infty$
and these are like $u \ra 0$.

We can see that 
the energy gap between the ground states and excited states
is order $\f{u^2}{R^2}$ where $R$ is a characteristic 
length scale of the submanifold which D$p$-brane 
is supposed to wrap.
Here we have chosen the vielbein as the induced metric and 
defined the $u$ as $T=u D$.
From this, we expect the parameter $\f{u}{R}$ 
controls the fuzziness and should be small if we want to have
the localized brane.
Moreover, 
in order to have the D$p$-brane 
we should take $1 \ll u$, which can be seen
from the boundary state in the path-integral formalism. 
Thus, we expect that 
we have the D$p$-brane wrapping the manifold
which locally looks like the flat D$p$-brane
if we can take 
\beq
1 \ll u \ll R,
\label{cond1}
\eeq
which implies $1 \ll R$ or $\sqrt{\alpha'} \ll R$ if we recover the $\alpha'$.
Note that we have defined $T$ and $u$ dimensionless.
We can not have such a D$p$-brane
if $R \sim \sqrt{\alpha'}$
because of the noncommutativity.
Let us consider a D2-brane on $S^2$ of radius 
$R \sim \sqrt{\alpha'}$ 
with unit flux as an example.
If we want to get the D2-brane localized on $S^2$,
it should have finite $u$ since 
it shrinks to the origin for $u=\infty$.
Thus we can not have the D2-brane localized on $S^2$
which is locally like the flat D2-brane.
One might think the D2-brane for $u \ra \infty$ is still 
described by the DBI action on $S^2$ of radius $R$.
However, 
the DBI action is not reliable, 
for example, for the D2-brane with finite $n$
on $S^2$
in the limit $u \ra \infty$ since the (higher derivative) corrections
to the DBI action can not be neglected.
In particular, there will be ${\cal O}(\f{u}{R})$ terms
in the action. 
Indeed, it should be described by the action of the $n$ BPS D0-branes
which are effectively on $S^2$ of radius $R_{eff}$.
Especially for $|n|=1$ and $n=0$,
the picture from the DBI action is completely wrong.

Here it should be emphasized that 
this conclusion does not mean
the usual analysis by the DBI action is wrong.
Indeed, it is well known that 
the DBI action is reliable 
only if higher derivative terms can be neglected, i.e.
$R/\sqrt{\alpha'} \gg 1$.
In this case, 
we can always take $u \ll R$ and 
the picture from the DBI action is indeed reliable.
As an example, let us consider the Myers effect \cite{Myers},
where the RR flux is turned on.
The analysis is reliable for $|n| \gg 1$, i.e. large magnetic flux. 
This means large $R$ by the on-shell condition
although we do not require on-shell condition in general.
On the other hand, it is also known that 
the DBI analysis is wrong if $|n|$ is finite.

Finally, 
we can think that the parameter $1/u$ 
represents the uncertainty of $\wh p$ \cite{Te}.
Thus the position is definite for $u \ra 0$, 
on the other hand, the position 
is ambiguous for $u \ra \infty$. 
This agrees with its interpretation as the parameter controlling
the noncommutativity of the world volume.

\subsection{Diffeomorphism and Seiberg-Witten Map}
\label{diffeoSW}

In this subsection, we will discuss
the diffeomorphism of the world volume of the D$p$-brane.
In the unstable D0-brane picture,
it is realized as a subgroup of 
the $U(N)$ gauge symmetry of the $N$ D0-branes \cite{AsSuTe1}.
This $U(N)$ gauge symmetry is 
unbroken if $T$, $\Phi^\mu$ and other fields are
proportional to the $N \times N$ identity matrix.
For the D$p$-brane configuration,
The $U(N)$ gauge symmetry is broken to the over all $U(1)$,
but a subgroup becomes gauge symmetry of the 
D$p$-brane, which contains 
the $U(1)$ gauge symmetry $e^{i \lambda(\wh x)}$, 
(the local Lorentz symmetry 
$e^{[\gamma^A,\gamma^B] c_{AB}} $) 
and the diffeomorphism 
$e^{i \eta^\alpha(\wh x) \wh p_\alpha}$
\cite{AsSuTe1, AsSuTe3}.\footnote{
This realization of the diffeomorphism 
is similar to the spectral action principle
\cite{spectral}.
If we formally consider 
$\bra{Dp} e^{-S_b} \ket{\! x \! = \! 0 \! } \sim {\rm Tr}(e^{-u^2 \wh H_0})$
and take $u \ll 1$, we have the spectral action without fermion
although the D-brane action is different from it.}

If we include $u$ into the definition of 
the vielbein $e^\mu_A \ra u e^\mu_A$, $u \ll 1$ corresponds to 
the large world volume metric 
$g_{\mu \nu}=e^A_\mu e^B_\nu \delta_{AB} \gg 1$.
As we have seen, $u \ll 1$ is a geometric region,
i.e. the world volume is commutative and localized.
The large metric expansion of the BSFT action
$S_{BSFT}=\f{2 \pi}{g_s} \bra{0} e^{-S_b} \ket{\! x \! = \! 0 \! }$ is \cite{AsSuTe1}
\beq
S_{BSFT} \sim \f{1}{g_s} \int dt \int dx^{p} \sqrt{g} \left( 1+2 \log 2 \, 
G_{\mu \nu} \f{\de f^\mu(x)}{\de x^\alpha} 
\f{\de f^\nu(x)}{\de x^\beta} g^{\alpha \beta} +{{\cal O}(g^{-2})}
\right)
\label{bsft1}
\eeq
where $G_{\mu \nu}=\delta_{\mu \nu}$
and $f^\mu(x)$ is the embedding of the world volume 
to the flat space.
This action is diffeomorphism invariant.
From the first term, we see that
the metric is related to the density of the 
unstable D-branes in this region of $u$
since the unstable D0-branes are almost independent each other
and the energy of the $N$ D0-branes is proportional to $N$.

On the other hand, for $u \gg 1$,
we should have the world volume action on the D$p$-branes,
for example, the DBI action for the flat world volume
with the constant $F$.
Actually, we know the equivalence between the boundary states.
It means that the $S_{BSFT}$ becomes the 
world volume action on the D$p$-branes
in the $u \ra \infty$ limit
because 
the disk partition function, $\f{2 \pi}{g_s} \langle {0} \ket{Dp}$,
gives the the effective action of the D$p$-brane
\cite{sigma}.
In leading order in $1/R$,
we will have the DBI action or the Nambu-Goto action
neglecting the gauge fields and other fields,
\beq
S_{BSFT} \sim \f{2 \pi}{g_s} \int dt \int dx^{p} 
\sqrt{{\rm det} \left(G_{\mu \nu} \f{\de f^\mu(x)}{\de x^\alpha} 
\f{\de f^\nu(x)}{\de x^\beta} \right)}.
\label{bsft2}
\eeq
We note that the volume factor in the (\ref{bsft2})
is much smaller than $\sqrt{g}$ in (\ref{bsft1})
since the induced metric is finite, but 
$g_{ij}$ is very large for $u \ll 1$.
We also note that (\ref{bsft2}) is valid 
only if (\ref{cond1}) is satisfied.
This is because there will be 
${\cal O}(\f{u}{R})$ terms 
and these terms can not be neglected
if (\ref{cond1}) is not satisfied.

Now let us consider 
a Hamiltonian $\wh H_0$ with a gap and take 
the $u \ra \infty$ limit.
Then the Hilbert space is reduced to 
the space spanned by the zero modes. 
We denote the projection operator to this space
as $P$, namely $P=\sum_{a=1}^{N_0} \ket{a} \bra{a}$.
Then by the gauge transformation $U \in U(N)$, which
includes the diffeomorphism, 
the projection operator becomes $P'=UPU^\dagger$.
Thus it gives another description of the system.
However, in $[P, U]=0$ case,
an operator $U'=PUP$ in the reduced space
is unitary and we can interpret this 
as a generator of the residual gauge symmetry.

Let us concentrate on the flat noncommutative D$(2p)$-brane.
In this case, there are a commutative 
and a noncommutative descriptions which are related by the
the Seiberg-Witten map \cite{SeWi}. 
Furthermore, 
the SW map can be almost understood as a diffeomorphism
\cite{Co, Is2, Ok1}.
Since the diffeomorphism is realized as the $U(N)$ gauge symmetry,
i.e. unitary transformation of the basis,
in our picture,
we expect that the Seiberg-Witten map is almost 
realized as the $U(N)$ gauge symmetry in our picture also.
Consider the D2-brane constructed from the unstable D0-branes with
\beq
T=u \gamma^\alpha (\wh p_\alpha 
+\f{1}{2} f_{\alpha \beta} \wh x^\beta
+ A_\alpha(\wh x) ), \,\,\,\,\,\,
\Phi^\alpha=\wh x^\alpha,
\label{aa1}
\eeq
where $f_{\alpha \beta}=- f_{\beta \alpha}$.
Then, $f_{\alpha \beta}$ is background flux and
$ A_\alpha(x)$ is the fluctuation of the gauge field 
around it.
The diffeomorphism discussed in \cite{Co, Is2, Ok1}
is given by
$U=e^{i \eta^\alpha(\wh x) \wh p_\alpha}$
such that $U^\dagger T U= 
u \gamma^A e^{\alpha}_A(\wh x)
(\wh p_\alpha + \f{i}{8} \omega_{BC,\alpha} [ \gamma^B, \gamma^C]  
+\f{1}{2} f_{\alpha \beta} \wh x^\beta )$ and
we define $\hat{A}_\alpha(x)$ as
$U^\dagger \Phi^\alpha U= \wh x^\alpha +(f^{-1})^{\alpha \beta}
\hat{A}_\beta(\wh x)$.
Note that there should be a map between $A_\alpha(x)$
and $\hat{A}_\alpha(x)$ up to the gauge transformation.
The vielbein $e^{\alpha}_A(\wh x)$
will be almost reduced to trivial one in the $u \ra \infty$ limit,
the D2-brane is almost described by
\beq
T=u \gamma^{\alpha}
(\wh p_\alpha 
+\f{1}{2} f_{\alpha \beta} \wh x^\beta ), \,\,\,\,\,\,
\Phi^\alpha = \wh x^\alpha +(f^{-1})^{\alpha \beta}
\hat{A}_\beta(\wh x) 
\label{aa2}
\eeq
in the $u \ra \infty$ limit.
Here the boundary states for these two configurations 
are actually those considered in \cite{Ok1}.
Thus
the map between $A_\alpha(x)$
and $\hat{A}_\alpha(x)$ is almost the SW map as shown in 
\cite{Ok1}.
In order to consider the obtain the exact SW map,
we need to have the zero modes of $T$ in (\ref{aa1})
and evaluate $(\tilde \Phi^\alpha )_{ab}=
\bra{a} \wh x^\alpha \ket{b}$.
Then the $\hat A_\alpha(x)$  should be given by
the formula
$\tilde \Phi^\alpha = \wh x^\alpha +(f^{-1})^{\alpha \beta}
\hat{A}_\beta(\wh x)$ where
$\wh x$ is the operator on the reduced Hilbert space
which is different from $\wh x$ in (\ref{aa1}).
This will give another representation of the SW map.
\footnote{
The SW map is not unique \cite{AsKi} and
indeed, the SW maps for 
the bosonic case and the superstring case are 
different \cite{OkTe, Ok}.
This difference might come from the 
subtlety of the UV regularization.
}


It is expected that 
the diffeomorphism realized as a subgroup of $U(N)$
in the $N$ unstable D0-branes
will be reduced to
the area preserving diffeomorphism realized as a subgroup 
of $U(N_0)$
in the $N_0$ BPS D0-branes after taking $u \ra \infty$ limit.
It is an interesting problem to explicitly show this.

\section{Conclusions}
\label{concl}

In this paper, 
we have studied the fuzzy or noncommutative D$p$-branes
in terms of infinitely many unstable D0-branes
and showed that the condensation of the tachyon 
of the unstable D0-branes
induces the noncommutativity.
We have also showed that
a boundary state for a D$p$-brane 
satisfying the Dirichlet boundary condition on a curved submanifold
embedded in the flat space
is not localized on the submanifold.
Although, we have considered a few examples,
it should be possible to generalize
to the D-brane wrapping $S^4$ \cite{Kara}, 
supertubes \cite{supertube} and others \cite{Hy}.

The $\f{1}{u^2} (\dot{x^\alpha})^2$ term (\ref{subi})
on the boundary of 
the world sheet 
can be regarded as 
the trace part of the massive symmetric tensor fields,
$W_{\alpha \beta}(x) \dot{x}^\alpha \dot{x}^\beta$.
It was argued that 
this trace part of the massive tensor 
can not be turned on
because it gives the nonzero one point-function
which corresponds to a linear term in the effective action
\cite{HaTe}.
However, from the D0-brane point of view,
this term corresponds to the tachyon 
and there is no apparent problem with this term,
especially, the effective action does not have 
a term linear in the tachyon.
This is valid 
at least 
if we regularize the Hilbert space by 
a finite dimensional space. 
The term is proportional to $1/u^2$ which means
that (\ref{subi}) is 
expanded by the inverse of the tachyon of the D0-branes.
Actually, the net effect of 
the problematic one-point function
in this case is a change of the normalization 
of the boundary state by a factor
$\exp (\f{\pi}{4 u^2})$ which can not be expanded
around $u=0$.
Hence it might be possible to think (\ref{subi})
as a background.
The role of this term may be interesting to study.

As mentioned in the introduction,
the noncommutative geometry of \cite{Connes} is represented
by the spectral triple which contains 
the Dirac operator or an analogue of it.
Because of it,
we can construct the analogue of the Riemannian geometry.
The 
Dirac operator corresponds to the tachyon operator
in our unstable D0-branes \cite{AsSuTe1}.
However, 
the noncommutative plane and the fuzzy sphere 
in the string theory \cite{CoDoSc,SeWi,Myers}
do not have such an object naturally. 
Indeed, as we have seen,
the information of the metric or the Dirac operator
will be almost dropped
in the $u \ra \infty$ limit, which
leads the noncommutative plane and the fuzzy sphere.
However, even in the $u \ra \infty$ limit,
the Dirac operator plays important role, for example,
in the computation of the
RR-charges which is directly related to
the Atiyah-Singer index theorem \cite{AsSuTe3}.
Moreover, if we can incorporate the space-time fermions,
the Dirac operator may play
important role as in \cite{spectral}.
From the point of view of the Connes' noncommutative geometry,
it may be interesting to study how
the commutative geometry with the nontrivial gauge field 
is related to the noncommutative geometry.
We hope to come back to this in the future.


\vskip6mm
\noindent
{\bf Acknowledgements}

\vskip2mm
The author would like to thank K.~Hashimoto, S.~Sugimoto
and W.~Taylor
for useful discussions.
This work was supported in part by DOE grant 
DE-FG02-96ER40949.
\\


\noindent

\appendix
\setcounter{equation}{0}

\section{Boundary state}
\label{bsr}

In this section we review the boundary states 
for the D-branes which are realized 
in the closed string Hilbert space.\footnote{
We omit the ghost part which does not play any role in this paper
because we do not consider the loop amplitudes.
We also omit the factor which depends on the time-direction,
which is universal for time-independent configuration.
See \cite{bdry} for those.}

The closed string operators at boundary of the world sheet are,
\begin{eqnarray}
X^\mu(\sigma)&=&\wh x_0^\mu+\sum_{m\ne 0}\frac{1}{\sqrt{|m|}}
(a_m^\mu e^{-im\sigma}+\wt a_m^\mu e^{im\sigma}), \\
P^\mu(\sigma)&=&\wh p_0^\mu-i\sum_{m\ne 0}{\sqrt{|m|}}
(a_m^\mu e^{-im\sigma}-\wt a_m^\mu e^{im\sigma}),
\end{eqnarray}
where
\begin{eqnarray}
a_{-m}^\mu=a_m^{\mu\dag}, \,\,
\wt a_{-m}^\mu=\wt a_m^{\mu\dag},~~~(m>0),
\end{eqnarray}
 which acts on the closed string Hilbert space.
The commutation relations are
\begin{eqnarray}
[X^\mu(\sigma),P^\nu(\sigma')]=i \delta^{\mu \nu} \delta(\sigma-\sigma'), 
\end{eqnarray}
\begin{eqnarray}
[a_m^{\mu},a_n^{\nu\dag}]=[ \wt a_m^{\mu}, \wt a_n^{\nu\dag} ]=
\delta_{m,n}\delta^{\mu \nu}.
\end{eqnarray}
The closed string Fock vacuum $\ket{0}$ is defined such that
\beq
\wh p_0^\mu \ket{0}=a_m^{\mu} \ket{0}=
\wt a_m^{\mu} \ket{0}=0 
\,\,\,\,\, (m >0).
\eeq

The coherent state of $X^\mu(\sigma)$ is defined as
\begin{eqnarray}
X^\mu(\sigma)\ket{x}&=&x^\mu(\sigma)\ket{x},
\end{eqnarray}
where
\begin{eqnarray}
x^\mu(\sigma)&=&x_0^\mu+\sum_{m\ne 0}\frac{1}{\sqrt{|m|}}\,
x_m^\mu e^{-im\sigma},~~~
\end{eqnarray}
which is explicitly written as
\beq
\ket{x}=\exp\left(
\sum_{m=1}^\infty \left(
-\f{1}{2} x_{-m} x_m -a_m^\dagger \wt a_m^{\dag}
+a_m^{\dag} x_m +x_{-m} \wt a_m^{\dag}
\right)
\right) \ket{x_0},
\eeq
where $\ket{x_0}$ satisfies 
$\wh x_0^\mu \ket{x_0}=x_0^\mu \ket{x_0}$ 
and $a_m^{\mu} \ket{x_0}=
\wt a_m^{\mu} \ket{x_0}=0 $.
Then the boundary state for a D0-brane at 
the origin is given as 
$\ket{D0} \equiv \ket{\! x \! = \! 0 \! }=\ket{x}|_{x_0=x_m=0}$ which satisfies 
\beq
X^\mu(\sigma) \ket{\! x \! = \! 0 \! }=0.
\eeq
It, of course, implies the Dirichlet boundary condition
$\dot{X}^\mu(\sigma) \ket{\! x \! = \! 0 \! }=0$.

The boundary states for D$p$-branes
is defined by
\begin{eqnarray}
\ket{Dp} \equiv \int[dx^\alpha]
\ket{x^\alpha, x^i=0} 
\label{Dpb}
\end{eqnarray}
where the superscript $\alpha=1, \ldots, p$ 
and $i=p+1, \ldots, 25$ represent the direction
tangent and transverse to the D-brane, respectively.
In fact, we can check that this satisfies 
the Neumann boundary condition,
$P_\alpha(\sigma) \ket{Dp}=0$.

The fields on the D-brane world-volume can be turned on through
a boundary interaction. 
Then,
the boundary state $\ket{Dp}$ are modified as
\begin{eqnarray}
\ket{Dp}_{S_b}&=&
e^{-S_b}\ket{Dp}
\label{BSbbs}
\end{eqnarray}
where $e^{-S_b}$ is an abbreviation 
for the boundary interaction:
\beq
e^{-S_b}=
\TrP \exp\left\{\oint
d\sigma \left( -S_b(X^\alpha(\sigma), P^i(\sigma)) \right)
\right\}.
\eeq
We can rewrite it as
\beq
e^{-S_b}\ket{Dp}=
\int[dx^\alpha]
\,
\TrP \exp\left\{\oint
d\sigma \left( -S_b(x^\alpha(\sigma), P^i(\sigma)) \right)
\right\}
\ket{x^\alpha, x^i=0}.
\eeq
Here 
$S_b(x^\alpha(\sigma), P^i(\sigma))$ 
is $N \times N$ matrix due to the Chan-Paton index
and $\TrP$ is the path-ordered trace for this index,
where $N$ is the number of the D$p$-branes we consider.
Note that if there is no boundary interaction, i.e. $S_b=0$,
we have $e^{-S_b}\ket{Dp}=N \ket{Dp}=\sum_{n=1}^N \ket{Dp}$.\footnote{
If we take $2 \pi T=-\log N$, the boundary state of the $N$ D$p$-branes
becomes
$N e^{-2 \pi T} \ket{Dp}=\ket{Dp}$. However, this is different
from one D$p$-brane which has different fluctuations on 
its world volume.
}

The boundary interaction for the gauge fields on the
D$p$-branes is given as \cite{Callan et al} 
\begin{eqnarray}
e^{-S_b}
=
\TrP \exp\left\{\oint
d\sigma\left(-A_\alpha(X)\dot X^\alpha
\right)\right\}.
\label{gauge2}
\end{eqnarray}
Similarly, the boundary interaction for the tachyons $T$
and the massless scalars $\Phi^i$ on the
D0-branes is given as 
\beq
e^{-S_b}
= \TrP \exp\left\{\oint
d\sigma\left( -T -i \Phi^i P_i
\right)\right\},
\label{D0bi}
\eeq
where $T$ and $\Phi^i$ are  $N \times N$ matrices
and $\sigma$ independent.
In particular, if we set $T=0$ and 
$\Phi^i={\rm diag} (a^i_1, \ldots, a^i_{N})$,
we have $e^{-S_b}\ket{D0}
=\sum_{n=1}^N e^{i \wh p_0^i a^i_n} \ket{D0}$,
i.e. $n$-th D0-brane is located at $x^i=a^i_n$.

Next we summarize the superstring case very briefly.
See \cite{AsSuTe1} for details.
In super string, the boundary states for D-branes
are obtained as a linear combinations of
the boundary states defined by
\begin{eqnarray}
\ket{Bp;\pm}=\int[dx^\alpha][d\psi^\alpha]
\ket{x^\alpha, x^i=0}\ket{\psi^\alpha, \psi^i=0;\pm},
\end{eqnarray}
where the superscript $\alpha$ and $i$ represent the direction
tangent and transverse to the D-brane, respectively.
The state $\ket{\theta}$ is the coherent state
for the world sheet fermions
satisfying
\beq
\label{coherent2}
\Psi_\pm^\mu(\sigma)\ket{\psi;\pm}=
\psi^\mu(\sigma)\ket{\psi;\pm},
\eeq
where
\beq
\Psi^\mu_\pm(\sigma)=
\Psi^\mu(\sigma) \pm i\wt\Psi^\mu(\sigma)
=\sum_{r}(\Psi^\mu_r e^{-ir\sigma}\pm i\wt\Psi^\mu_r e^{ir\sigma}),
\,\,\,\; \psi^\mu(\sigma)
=\sum_{r}\psi_r^\mu e^{-ir\sigma}.
\eeq
In this paper, we will omit the subscript $\pm$
because it does not play any role in this paper.
We also introduce 
$\Pi_\mu(\sigma)$ which is the conjugate momentum
of $\Psi^\mu(\sigma)$. 

For simplicity, we consider $N$ $\dz$ pairs in type IIA and set 
$\Phi^\mu=\tilde \Phi^\mu$ 
where $N \times N$ matrices 
$\Phi$ and $\tilde \Phi$ are massless scalars on D0 and
$\overline{\mbox{D}0}$, respectively.
We define a $2N \times 2 N$ matrix 
$\mat{\Phi^\mu,0,0,\tilde \Phi^\mu}
=\Phi^\mu 1_{2 \times 2}$ and we will denote it
as $\Phi^\mu$ in this paper.
We also denote a $2N \times 2N$ Hermitian matrix 
$\mat{0,T,T^\dagger,0}$
as $T$ for notational simplicity.
Then the boundary interaction is \cite{TaTeUe} \cite{KrLa} 
\beq
e^{-S_b}= 
\TrP \exp \int d \sigma \left( 
-i \Phi^\mu P_\mu(\sigma) 
-\f{1}{2} [\Phi^\mu, \Phi^\nu] \Pi_\mu(\sigma)\Pi_\nu(\sigma)
-T^2 + \Pi_\mu(\sigma) [T,\Phi^\mu] 
\right),
\label{D0D0bar}
\eeq
where 
$\TrP$ is the path-ordered trace for the $2N \times 2N$ matrix. 
From this, we also see that 
boundary interaction for the $N_0$ BPS D0-branes is
\beq
e^{-S_b}= 
\TrP \exp \int d \sigma \left( 
-i \Phi^\mu P_\mu(\sigma) 
-\f{1}{2} [\Phi^\mu, \Phi^\nu] \Pi_\mu(\sigma)\Pi_\nu(\sigma)
\right),
\label{BPSD0}
\eeq
where  $\TrP$ is the path-ordered trace for 
the $N_0 \times N_0$ matrix.


\section{Normalization of the boundary state}
\label{normal}

In this appendix, we discuss about the normalization 
of the boundary state.

First we introduce 
a normalization factor $A$ to the path-integral measure
in (\ref{hl}) as $\TrP e^{-i \int d\sigma H}=A \int [d x]e^{i \int d \sigma
L}$,
where 
\beq
A=\f{\bra{0} \TrP e^{-i \int d\sigma \wh H} \ket{D0}}
{\bra{ 0} \int [dx] e^{i \int d\sigma L} \ket{D0}}.
\label{defa}
\eeq
Formally we can rewire it as 
$A=\Tr e^{-u^2 \wh H_0}/\int [d x(\sigma)] e^{-\int d\sigma L_0}$.
Then in (\ref{hl}) we can use a path-integral measure
with any normalization.
Especially, we can use the measure for the perturbative theory.

Now let us recall that 
the BSFT action $S_{BSFT}$ 
(for on-shell fields) is proportional to
the disk partition function \cite{BSFT, KMM1, GeSh, 
KMM2, KrLa, TaTeUe}, 
and the normalization 
of the BSFT action should be fixed such that
it correctly reproduce the tension of the D$p$-brane.
Using the boundary state, 
we can define $S_{BSFT}=\f{2 \pi}{g_s} \bra{0} Dp \rangle$
by appropriately 
choosing the normalization of the path-integral
measure in (\ref{Dpb}).
For the D$(-1)$-brane, we have used 
$\bra{0} D(-1) \rangle=\bra{0} x=0 \rangle=
\bra{0} e^{-\sum a^\dagger \tilde a^\dagger} \ket{x_0}=1$.
Therefore, with this normalization,
$A$ should be $1$ 
if (\ref{hl}) or its supersting analogue
gives the correct tension of the D$p$-brane.
This condition is equivalent to the 
condition that
the D$p$-brane solution in the $N$ D0-branes
has correct tension.
This was indeed shown in \cite{Te}
and the correct ration is obtained for superstring.
Furthermore, if we use the zeta-function regularization
for the path-integral measure \cite{AsSuTe1, AsSuTe3},
we have $A=1$ though it is rather formal.
In this paper, we will use a definition of 
the path-integral measure $[dx]$ (or its supersymmetric 
extension) such that $A=1$.

For the bosonic case, 
we expect $A=1$ since the correct tension is obtained
in the decent relation \cite{KMM1}.
Note that the normalization of the boundary state is related 
to the constant part of the tachyon $t(x)$ 
and
we should take $t(x)$ such that $T \geq 0$ for bosonic case, 
i.e. the eigen values of the $T$ are zero or positive.
(At least, there should be no eigen value which 
goes to $-\infty$.)
This is because the tachyon potential $V(T)$ goes 
to $-\infty$ as $T \rightarrow -\infty$ and 
the negative tachyon region may be forbidden
although we do not know a correct interpretation of it.

Finally, we note that 
the disk partition function used in (\ref{defa}) has
the UV divergence of the world sheet and we should regularize
it as usual. 
This can be done by replacing 
$P(\sigma)$ by $e^{\epsilon \f{\de^2}{\de \sigma^2}} P(\sigma)$ which is
equivalent to the regularization used in \cite{sigma}.
Then the fluctuations $A_\alpha(\wh x), \phi^i(\wh x)$ 
and $t(\wh x)$ will depend on 
the regularization parameter
$\epsilon$.

\section{Decoupling of the D0-branes}
\label{hs}

In this appendix
we will first consider the validity of 
the assumption leading (\ref{bp2}) or (\ref{bp2a})
and then we will consider the Harmonic oscillator case
which violates the assumption.

Let us consider the case in bosonic string
with $t(x)=0$.
Here $\wh H=u^2 H_0 +i f^\mu(x) P_\mu(\sigma)$.
Then we can regard $u^2 H_0$ as the ``free Hamiltonian''
and $H_I=i f^\mu(x) P_\mu(\sigma)$ as the interaction Hamiltonian.
As usual 
in the interaction picture, we expand $e^{\-S_b}$ as
\beq
e^{-S_b}=\sum_{n=0}^\infty
\Tr
\left( 
e^{-2 \pi u^2 H_0} 
\prod_{l=1}^n \int_0^{2\pi} d\sigma_l 
\Theta(\sigma_{l}-\sigma_{l-1})
\,\,
H_I(\sigma_l) 
\right)
\eeq
where $H_I(\sigma)=i e^{\sigma u^2 H_0} 
 f^\mu(x) 
e^{-\sigma u^2 H_0} P_\mu(\sigma)$
and $\sigma_0=0$.
Thus we can evaluate it using the eigen states of $H_0$
as in \cite{El}.
It is easy to see that
the most dangerous terms comes from 
the integrations near $\sigma_l=\sigma_{l-1}$
and they are ${\cal O} (\f{1}{u^2})$.
Thus if the summation over the energies
is finite, the assumption is valid
at least in this perturbation theory.
Actually, for the circle or sphere D$p$-brane case
we can easily see that 
only a finite number of the energy eigen states contributes
because of the rotation symmetry.
By adding some smooth terms to $f^\mu(x)$,
we can consider the D$p$-brane of generic shape
and the summation over the energies
is finite.
Thus we expect that 
the assumption is valid for the D$p$-brane with $t(x)=0$.

Let us consider the Hamiltonian of the Harmonic oscillator
in the bosonic string.
We take
\begin{equation}
u^2 t=\f{\alpha^2}{2} \wh x^2 - \f{1}{2} u \alpha.
\end{equation}
Then 
\beq
u^2 \wh H_0= \f{u^2}{2} \wh p^2+ \f{\alpha^2}{2} \wh x^2 - \f{1}{2} u \alpha
\eeq
is a harmonic oscillator
with a mass $\s{m} \omega=\alpha$ and
a frequency $\f{1}{\s{m}}=u$.
The eigen state $\ket{n}$ $(n \in Z \geq 0)$ has 
the energy $E_n=u \alpha n$.
The expectation value of the position operator
for the ground state vanishes, $\bra{0} \wh x \ket{0}=0$.
in the $u \ra \infty$ limit one might think
only the ground state contributes in the $\TrP$
as argued in section \ref{red} and 
obtain 
\begin{equation}
\ket{B}= \TrP e^{-i\int d \sigma \hat{H}(\hat{x},\hat{p}) }\ket{\! x \! = \! 0 \! }
= \ket{\! x \! = \! 0 \! }
\end{equation}
in the $u \ra \infty$ limit.
However, this observation is not correct because
the excited states do contribute in the $\TrP$
even in the $u \ra \infty$ limit.
Indeed, at the second order perturbation theory,
$  \TrP e^{-i\int d \sigma \hat{H}(\hat{x},\hat{p}) }$ is 
\begin{eqnarray}
&& \sum_{n=0}^\infty \int_0^{2\pi} d\sigma_1 \int_0^{2\pi} d\sigma_2
\Theta(\sigma_2-\sigma_1) 
\bra{n} e^{-(2\pi-\sigma_2) T} (i\wh x P(\sigma_2) ) e^{-(\sigma_2-\sigma_1)T}
 (i\wh x P(\sigma_1) ) e^{-\sigma_1 T} \ket{n} \nonumber \\
&=& \sum_{n,m=0}^\infty
\int_0^{2\pi} d\sigma_2 \int_0^{\sigma_2} d\sigma_1 
e^{-2\pi n u \alpha} e^{ (\sigma_2-\sigma_1)(n-m) u \alpha}
\sum_{n=0}^\infty \bra{n}  (i\wh x) \ket{m} \bra{m }  (i\wh x) \ket{n}
P(\sigma_1)P(\sigma_2), \nonumber
\end{eqnarray}
and if we set $P(\sigma)$
to its zero mode $p_0$ for simplicity\footnote{
This is formally archived by considering
$\bra{D1}e^{-i p_0 \wh x_0} e^{-S_b} \ket{\! x \! = \! 0 \! }$.}
and take the $u \ra \infty$ limit which implies $n=0$,
it is reduced to
\begin{eqnarray}
&& \sum_m^\infty \int_0^{2\pi} d\sigma_2 
\frac{1}{u \alpha m} (1-e^{ -\sigma_2 m u \alpha})
\bra{0} i \wh x \ket{m} \bra{m} i \wh x \ket{0} p_0^2 \CR
&=& -\frac{2 \pi}{u \alpha} |\bra{0} \wh x \ket{1}|^2 p_0^2
+\left({\rm higher \, order \, in} \, \f{1}{u}\right).
\end{eqnarray}
Here, the integrations near $\sigma_1=\sigma_2$ 
gave the leading term.
We can compute $|\bra{0} \wh x \ket{1}|^2=\frac{u}{\alpha}$
using a following transformation
\begin{eqnarray}
 \wh p=\sqrt{\frac{\alpha}{u}} \wh p', \,\,
 \wh x=\sqrt{\frac{u}{\alpha}} \wh x',
\end{eqnarray}
which keeps the commutation relation invariant 
$[\wh x', \wh p']=i$ and 
$T=u \alpha (\frac{\wh p'^2}{2}+ \frac{\wh x'^2}{2}-\f{1}{2})$.
Therefore, it becomes
\begin{equation}
  -\frac{2 \pi}{u \alpha} |\bra{0} \wh x \ket{1}|^2 p_0^2=
-2 \pi \frac{p_0^2}{\alpha^2}, 
\end{equation}
which is indeed finite in the  $u \ra \infty$ limit.
This means that 
the infinitely many unstable D$0$-branes 
are not reduced to a D$0$-brane represented by the 
ground state $\ket{0}$ and 
the system is not localized at the point
in the $u \ra \infty$ limit.

Actually,
the boundary interaction is written as
\begin{eqnarray}
&& \TrP e^{-\int d \sigma 
\left( u \alpha 
\left(\frac{\wh p'^2}{2}+ \frac{\wh x'^2}{2}-1  \right) 
- i \wh x' \sqrt{u}{\alpha} P(\sigma)
\right)}, \CR
&=& \TrP e^{-\int d \sigma 
\left( u \alpha 
\left(\frac{\wh p'^2}{2}+ 
\frac{(\wh x'-i\sqrt{\frac{1}{u \alpha^3}}P(\sigma))^2}{2}-1  \right) 
- \frac{1}{2 \alpha^2} P(\sigma)^2
\right)}.
\end{eqnarray}
If we take $P(\sigma)$
to its zero mode $p_0$,
it is reduced to
\begin{eqnarray}
e^{- \frac{1}{2 \alpha^2} p_0^2},
\label{zerox}
\end{eqnarray}
in the $u \ra \infty$ limit
and its Fourier transform is 
$\alpha e^{- \frac{\alpha^2}{2}  x_0^2}$.
Thus the result is an object smeared in the 
region $|x| < \frac{1}{\alpha}$.

In the path-integral formalism,
we have 
\begin{eqnarray}
 - L_0= -\frac{1}{2 u^2} \dot{x}(\sigma)^2 -
 \frac{\alpha^2}{2 }x(\sigma)^2 +\f{1}{2} u \alpha
 \ra_{u \ra \infty} -\frac{\alpha^2}{2} x(\sigma)^2 
+\f{1}{2} u \alpha,
\end{eqnarray}
which is the tachyon profile studied in 
the bosonic BSFT \cite{KMM1} 
and was interpreted as the
tachyon kink solution in the D$1$-brane, which represents 
a D$0$-brane in the limit $\alpha \ra \infty$.
We note that 
the zero  mode sector of it is 
exactly coincide with the (\ref{zerox}).

We note that 
instead of D0-branes in the bosonic string theory
we can repeat the above consideration for 
the non BPS D0-branes in type IIB string theory.
In this case, there is no constant shift like $u \alpha/2$.

\newpage

\newcommand{\J}[4]{{\sl #1} {\bf #2} (#3) #4}
\newcommand{\andJ}[3]{{\bf #1} (#2) #3}
\newcommand{\AP}{Ann.\ Phys.\ (N.Y.)}
\newcommand{\MPL}{Mod.\ Phys.\ Lett.}
\newcommand{\NP}{Nucl.\ Phys.}
\newcommand{\PL}{Phys.\ Lett.}
\newcommand{\PR}{ Phys.\ Rev.}
\newcommand{\PRL}{Phys.\ Rev.\ Lett.}
\newcommand{\PTP}{Prog.\ Theor.\ Phys.}
\newcommand{\hep}[1]{{\tt hep-th/{#1}}}


\begin{thebibliography}{99}
\baselineskip=12pt


\bibitem{CoDoSc}
  A.~Connes, M.~R.~Douglas and A.~Schwarz,
  ``Noncommutative geometry and matrix theory: Compactification on tori,''
  JHEP {\bf 9802} (1998) 003
  [arXiv:hep-th/9711162].


\bibitem{DoHu}
  M.~R.~Douglas and C.~M.~Hull,
  ``D-branes and the noncommutative torus,''
  JHEP {\bf 9802} (1998) 008
  [arXiv:hep-th/9711165].


\bibitem{SeWi}
N.~Seiberg and E.~Witten,
``String theory and noncommutative geometry,''
JHEP {\bf 9909} (1999) 032,
hep-th/9908142.

\bibitem{Myers}
R.C. Myers,
``Dielectric-Branes,''
JHEP {\bf 9912} (1999) 022,
hep-th/9910053.


\bibitem{Connes}
A. Connes,
``Noncommutative geometry,'' Academic Press, 1994.\\
See also,\\
A. Connes, ``A Short Survey of Noncommutative Geometry,''
hep-th/0003006.\\
A. Connes, ``Noncommutative Geometry Year 2000,''
math.QA/0011193.




\bibitem{AsSuTe1}
  T.~Asakawa, S.~Sugimoto and S.~Terashima,
  ``D-branes, matrix theory and K-homology,''
  JHEP {\bf 0203} (2002) 034
  [hep-th/0108085];

\bibitem{AsSuTe2}
  T.~Asakawa, S.~Sugimoto and S.~Terashima,
  ``D-branes and KK-theory in type I string theory,''
  JHEP {\bf 0205} (2002) 007
  [hep-th/0202165];


\bibitem{AsSuTe3}
  T.~Asakawa, S.~Sugimoto and S.~Terashima,
  ``Exact description of D-branes via tachyon condensation,''
  JHEP {\bf 0302} (2003) 011
  [hep-th/0212188];
  ``Exact description of D-branes in K-matrix theory,''
  Prog.\ Theor.\ Phys.\ Suppl.\  {\bf 152} (2004) 93
  [hep-th/0305006];



\bibitem{Te}
S.~Terashima, ``A construction of commutative D-branes from lower
dimensional non-BPS D-branes,'' JHEP {\bf 0105}, 059 (2001)
[arXiv:hep-th/0101087].


\bibitem{TaTe}
  T.~Takayanagi and S.~Terashima,
  ``c = 1 matrix model from string field theory,''
  hep-th/0503184.


\bibitem{BFSS}
T.~Banks, W.~Fischler, S.~H.~Shenker and L.~Susskind, ``M theory
as a matrix model: A conjecture,'' Phys.\ Rev.\ D {\bf 55} (1997)
5112 [hep-th/9610043].


\bibitem{deWit}
  B.~de Wit, J.~Hoppe and H.~Nicolai,
  ``On The Quantum Mechanics Of Supermembranes,''
  Nucl.\ Phys.\ B {\bf 305} (1988) 545.









\bibitem{KMM1}
D.~Kutasov, M.~Marino and G.~W.~Moore, ``Some exact results on
tachyon condensation in string field theory,'' JHEP {\bf 0010}
(2000) 045 [arXiv:hep-th/0009148].


\bibitem{GeSh}
  A.~A.~Gerasimov and S.~L.~Shatashvili,
  ``On exact tachyon potential in open string field theory,''
  JHEP {\bf 0010} (2000) 034
  [arXiv:hep-th/0009103].


\bibitem{KMM2}
D.~Kutasov, M.~Marino and G.~W.~Moore, ``Remarks on tachyon
condensation in superstring field theory,'' arXiv:hep-th/0010108.

\bibitem{KrLa}
P.~Kraus and F.~Larsen, ``Boundary string field theory of the
DD-bar system,'' Phys.\ Rev.\ D {\bf 63} (2001) 106004
[arXiv:hep-th/0012198].


\bibitem{TaTeUe}
T.~Takayanagi, S.~Terashima and T.~Uesugi, ``Brane-antibrane
action from boundary string field theory,'' JHEP {\bf 0103} (2001)
019 [arXiv:hep-th/0012210].

\bibitem{HaTe3}
  K.~Hashimoto and S.~Terashima,
  ``Stringy derivation of Nahm construction of monopoles,''
  hep-th/0507078.

\bibitem{Is}
N. Ishibashi,
``$p$-branes from $(p-2)$-branes
in the Bosonic String Theory,''
Nucl. Phys. {\bf B539} (1999) 107,
hep-th/9804163.

\bibitem{El}
  I.~Ellwood,
  ``Relating branes and matrices,''
  hep-th/0501086.


\bibitem{Re}
  V.~M.~Red'kov,
  ``Generally relativistical Tetrode-Weyl-Fock-Ivanenko formalism and
  behaviour of quantum-mechanical particles of spin 1/2 in the Abelian
  monopole field,''
  arXiv:quant-ph/9812002.

\bibitem{Ab}
  A.~A.~.~Abrikosov,
  ``Dirac operator on the Riemann sphere,''
  arXiv:hep-th/0212134. 



\bibitem{Haldane}
  F.~D.~M.~Haldane,
  ``Fractional Quantization Of The Hall Effect: A Hierarchy Of Incompressible
  Phys.\ Rev.\ Lett.\  {\bf 51} (1983) 605.


\bibitem{KaTa}
  D.~Kabat and W.~I.~Taylor,
  ``Spherical membranes in matrix theory,''
  Adv.\ Theor.\ Math.\ Phys.\  {\bf 2} (1998) 181
  [arXiv:hep-th/9711078];
  W.~I.~Taylor,
  ``Lectures on D-branes, gauge theory and M(atrices),''
  arXiv:hep-th/9801182.


\bibitem{HiNoSu}
  Y.~Hikida, M.~Nozaki and Y.~Sugawara,
  ``Formation of spherical D2-brane from multiple D0-branes,''
  Nucl.\ Phys.\ B {\bf 617} (2001) 117
  [arXiv:hep-th/0101211].


\bibitem{roll}
A.~Sen, ``Rolling tachyon,'' JHEP {\bf 0204} (2002) 048
[arXiv:hep-th/0203211];
``Tachyon matter,'' JHEP {\bf 0207} (2002) 065
[arXiv:hep-th/0203265].


\bibitem{Ha}
  K.~Hashimoto,
  ``The shape of non-Abelian D-branes,''
  JHEP {\bf 0404} (2004) 004
  [arXiv:hep-th/0401043].




\bibitem{HaTe}
K.~Hashimoto and S.~Terashima, ``Boundary string field theory as a
field theory: Mass spectrum and interaction,'' JHEP {\bf 0410}
(2004) 040, hep-th/0408094.


\bibitem{TaRa}
  W.~I.~Taylor and M.~Van Raamsdonk,
  ``Multiple D0-branes in weakly curved backgrounds,''
  Nucl.\ Phys.\ B {\bf 558} (1999) 63
  [arXiv:hep-th/9904095];
  ``Multiple Dp-branes in weak background fields,''
  Nucl.\ Phys.\ B {\bf 573} (2000) 703
  [arXiv:hep-th/9910052].


\bibitem{Sa}
  Y.~Sato,
  ``Dissolving D0-brane into D2-brane with background B-field,''
  arXiv:hep-th/0505045.







\bibitem{spectral}
  A.~H.~Chamseddine and A.~Connes,
  ``The spectral action principle,''
  Commun.\ Math.\ Phys.\  {\bf 186} (1997) 731
  [arXiv:hep-th/9606001].


\bibitem{sigma}
  E.~S.~Fradkin and A.~A.~Tseytlin,
  ``Nonlinear Electrodynamics From Quantized Strings,''
  Phys.\ Lett.\ B {\bf 163} (1985) 123;
O.~D.~Andreev and A.~A.~Tseytlin,
 ``Partition Function Representation For The Open Superstring
 Effective Action:
Cancellation Of Mobius Infinities And
Derivative Corrections To Born-Infeld
Lagrangian,'' Nucl.\ Phys.\ B {\bf 311} (1988) 205.


\bibitem{Is2}
N.~Ishibashi,
``A relation between commutative and noncommutative
descriptions of  D-branes,''
hep-th/9909176.

\bibitem{Co}
L.~Cornalba,
``D-brane physics and noncommutative Yang-Mills theory,''
Adv.\ Theor.\ Math.\ Phys.\  {\bf 4} (2000) 271,
hep-th/9909081.


\bibitem{Ok1}
K. Okuyama,
``A Path Integral Representation of the Map between
Commutative and Noncommutative Gauge Fields,''
JHEP {\bf 0003} (2000) 016,
hep-th/9910138.


\bibitem{Kara}
  J.~Castelino, S.~M.~Lee and W.~I.~Taylor,
  Nucl.\ Phys.\ B {\bf 526} (1998) 334
  [arXiv:hep-th/9712105];
  D.~Karabali, V.~P.~Nair and S.~Randjbar-Daemi,
  ``Fuzzy spaces, the M(atrix) model and the quantum Hall effect,''
  arXiv:hep-th/0407007.


\bibitem{supertube}
  D.~Mateos and P.~K.~Townsend,
  Phys.\ Rev.\ Lett.\  {\bf 87} (2001) 011602
  [arXiv:hep-th/0103030].

\bibitem{Hy}
  Y.~Hyakutake,
  ``Torus-like dielectric D2-brane,''
  JHEP {\bf 0105} (2001) 013
  [arXiv:hep-th/0103146].

\bibitem{bdry}
For a review,
P. Di Vecchia and A. Liccardo,
``D-branes in string theory, I,''
hep-th/9912161.


\bibitem{Callan et al}
C.~G.~Callan, C.~Lovelace, C.~R.~Nappi and S.~A.~Yost,
``Adding Holes And Crosscaps To The Superstring,''
Nucl.\ Phys.\ B {\bf 293} (1987) 83;
``Loop Corrections to Superstring Equations of Motion,''
Nucl.\ Phys.\ B {\bf 308} (1988) 221.


\bibitem{BSFT}
E.~Witten, ``On background independent open string field theory,''
Phys.\ Rev.\ D {\bf 46} (1992) 5467 [arXiv:hep-th/9208027];
``Some Computations in Background Independent Open-String Field
Theory,'' Phys.\ Rev.\ D {\bf 47} (1993) 3405
[arXiv:hep-th/9210065];

S.~L.~Shatashvili, ``Comment on the background independent open
string theory,'' Phys.\ Lett.\ B {\bf 311} (1993) 83;
[arXiv:hep-th/9303143];
``On the problems with background independence in string theory,''
Alg.\ Anal.\  {\bf 6}, 215 (1994) [arXiv:hep-th/9311177].



\bibitem{Senreview}
A.~Sen, ``Tachyon dynamics in open string theory,''
hep-th/0410103.







\bibitem{AsKi}
  T.~Asakawa and I.~Kishimoto,
  ``Comments on gauge equivalence in noncommutative geometry,''
  JHEP {\bf 9911} (1999) 024
  [arXiv:hep-th/9909139].

\bibitem{OkTe}
  Y.~Okawa and S.~Terashima,
  ``Constraints on effective Lagrangian of D-branes from non-commutative  gauge
  theory,''
  Nucl.\ Phys.\ B {\bf 584} (2000) 329
  [arXiv:hep-th/0002194].

\bibitem{Ok}
  Y.~Okawa,
  ``Derivative corrections to Dirac-Born-Infeld Lagrangian and  non-commutative
  gauge theory,''
  Nucl.\ Phys.\ B {\bf 566} (2000) 348
  [arXiv:hep-th/9909132];
  S.~Terashima,
  ``On the equivalence between noncommutative and ordinary gauge theories,''
  JHEP {\bf 0002} (2000) 029
  [arXiv:hep-th/0001111];
  ``The non-Abelian Born-Infeld action and noncommutative gauge theory,''
  JHEP {\bf 0007} (2000) 033
  [arXiv:hep-th/0006058].






\bibitem{Sen0}
A.~Sen,
``$SO(32)$ Spinors of Type I and Other Solitons
on Brane-Antibrane Pair,''
JHEP {\bf 9809} (1998) 023,
hep-th/9808141;\\
A.~Sen,
``Non-BPS states and branes in string theory,''
hep-th/9904207, and references therein.

\bibitem{Halpern}
K. Bardakci, ``Dual Models and Spontaneous Symmetry Breaking,''
Nucl. Phys. {\bf B68} (1974) 331;
K. Bardakci and M. B. Halpern, ``Explicit Spontaneous Breakdown
in a Dual Model,''
Phys. Rev. {\bf D10} (1974) 4230;
K. Bardakci and M. B. Halpern, ``Explicit Spontaneous Breakdown
in a Dual Model II: N Point Functions,''
Nucl. Phys. {\bf B96} (1975) 285;
K. Bardakci, ``Spontaneous Symmetry Breakdown in the Standard Dual
String Model,''
Nucl. Phys. {\bf B133} (1978) 297;























































\end{thebibliography}
\end{document}